\documentclass[twocolumn]{aastex62}
\graphicspath{{./}{figures/}}
\usepackage{verbatim}
\shorttitle{Solar Wind Sounding}
\shortauthors{Madison et al.}
\begin{document}
\title{THE NANOGRAV 11-YEAR DATA SET: SOLAR WIND SOUNDING THROUGH PULSAR TIMING}
\correspondingauthor{D. R. Madison}
\email{dmadison@nrao.edu}

\author[0000-0003-2285-0404]{D. R. Madison}
\affil{The National Radio Astronomy Observatory, 520 Edgemont Rd., Charlottesville, VA, 22903, USA}

\author{J. M. Cordes}
\affil{Department of Astronomy, Cornell University, 616-A Space Sciences Building, Ithaca, NY 14853, USA}
\affil{Cornell Center for Astrophysics and Planetary Science, 104 Space Sciences Building, Ithaca, NY 14853, USA}

\author{Z. Arzoumanian}
\affil{Center for Research and Exploration in Space Science and Technology and X-Ray Astrophysics Laboratory, NASA Goddard Space Flight Center, Code 662, Greenbelt, MD 20771, USA}

\author{S. Chatterjee}
\affil{Department of Astronomy, Cornell University, 616-A Space Sciences Building, Ithaca, NY 14853, USA}
\affil{Cornell Center for Astrophysics and Planetary Science, 104 Space Sciences Building, Ithaca, NY 14853, USA}

\author{K. Crowter}
\affil{Department of Physics and Astronomy, University of British Columbia, 6224 Agricultural Rd., Vancouver, BC V6T 1Z1, Canada}

\author{M. E. DeCesar}
\affil{Department of Physics, Lafayette College, Easton, PA 18042, USA}

\author{P. B. Demorest}
\affil{The National Radio Astronomy Observatory, 1003 Lopezville Rd., Socorro, NM 87801, USA}

\author[0000-0001-8885-6388]{T. Dolch}
\affil{Department of Physics, Hillsdale College, 33 E. College St., Hillsdale, MI 49242, USA}

\author{J. A. Ellis}
\affil{Department of Physics and Astronomy, West Virginia University, P.O. Box 6315, Morgantown, WV 26506, USA}
\affil{Center for Gravitational Waves and Cosmology, West Virginia University, Chestnut Ridge Research Building, Morgantown, WV 26505, USA}

\author{R. D. Ferdman}
\affil{School of Chemistry, University of East Anglia, Norwich, NR4 TTJ, United Kingdom}

\author{E. C. Ferrara}
\affil{NASA Goddard Space Flight Center, Greenbelt, MD 20771, USA}

\author[0000-0001-8384-5049]{E. Fonseca}
\affil{Department of Physics, McGill University, 3600 University St., Montreal, QC H3A 2T8, Canada}

\author{P. A. Gentile}
\affil{Department of Physics and Astronomy, West Virginia University, P.O. Box 6315, Morgantown, WV 26506, USA}
\affil{Center for Gravitational Waves and Cosmology, West Virginia University, Chestnut Ridge Research Building, Morgantown, WV 26505, USA}

\author{G. Jones}
\affil{Department of Physics, Columbia University, New York, NY 10027, USA}

\author{M. L. Jones}
\affil{Department of Physics and Astronomy, West Virginia University, P.O. Box 6315, Morgantown, WV 26506, USA}
\affil{Center for Gravitational Waves and Cosmology, West Virginia University, Chestnut Ridge Research Building, Morgantown, WV 26505, USA}

\author[0000-0003-0721-651X]{M. T. Lam}
\affil{Department of Physics and Astronomy, West Virginia University, P.O. Box 6315, Morgantown, WV 26506, USA}
\affil{Center for Gravitational Waves and Cosmology, West Virginia University, Chestnut Ridge Research Building, Morgantown, WV 26505, USA}

\author{L. Levin}
\affil{Department of Physics and Astronomy, West Virginia University, P.O. Box 6315, Morgantown, WV 26506, USA}
\affil{Center for Gravitational Waves and Cosmology, West Virginia University, Chestnut Ridge Research Building, Morgantown, WV 26505, USA}

\author{D. R. Lorimer}
\affil{Department of Physics and Astronomy, West Virginia University, P.O. Box 6315, Morgantown, WV 26506, USA}
\affil{Center for Gravitational Waves and Cosmology, West Virginia University, Chestnut Ridge Research Building, Morgantown, WV 26505, USA}

\author{R. S. Lynch}
\affil{Green Bank Observatory, P.O. Box 2, Green Bank, WV 24944, USA}

\author[0000-0001-7697-7422]{M. A. McLaughlin}
\affil{Department of Physics and Astronomy, West Virginia University, P.O. Box 6315, Morgantown, WV 26506, USA}
\affil{Center for Gravitational Waves and Cosmology, West Virginia University, Chestnut Ridge Research Building, Morgantown, WV 26505, USA}

\author{C. M. F. Mingarelli}
\affil{Center for Computational Astrophysics, Flatiron Institute, 162 5th Ave, New York, NY 10010, USA}

\author[0000-0002-3616-5160]{C. Ng}
\affil{Department of Physics and Astronomy, University of British Columbia, 6224 Agricultural Rd., Vancouver, BC V6T 1Z1, Canada}
\affil{Dunlap Institute for Astronomy and Astrophysics, University of Toronto, 50 St. George St., Toronto, ON M5S 3H4, Canada}

\author[0000-0002-6709-2566]{D. J. Nice}
\affil{Department of Physics, Lafayette College, Easton, PA 18042, USA}

\author[0000-0001-5465-2889]{T. T. Pennucci}
\affil{Institute of Physics, E\"{o}tv\"{o}s Lor\'{a}nd University, P\'{a}zm\'{a}ny P.s. 1/A, 1117 Budapest, Hungary}

\author[0000-0001-5799-9714]{S. M. Ransom}
\affil{The National Radio Astronomy Observatory, 520 Edgemont Rd., Charlottesville, VA, 22903, USA}

\author{P. S. Ray}
\affil{Space Science Division, Naval Research Laboratory, Washington, DC 20375-5352, USA}

\author[0000-0002-6730-3298]{R. Spiewak}
\affil{Centre for Astrophysics and Supercomputing, Swinburne University of Technology, P.O. Box 218, Hawthorn, Victoria 3122, Australia}
\affil{Center for Gravitation, Cosmology and Astrophysics, Department of Physics, University of Wisconsin-Milwaukee, P.O. Box 413, Milwaukee, WI 53201, USA}

\author[0000-0001-9784-8670]{I. H. Stairs}
\affil{Department of Physics and Astronomy, University of British Columbia, 6224 Agricultural Rd., Vancouver, BC V6T 1Z1, Canada}

\author{K. Stovall}
\affil{The National Radio Astronomy Observatory, 1003 Lopezville Rd., Socorro, NM 87801, USA}

\author{J. K. Swiggum}
\affil{Center for Gravitation, Cosmology and Astrophysics, Department of Physics, University of Wisconsin-Milwaukee, P.O. Box 413, Milwaukee, WI 53201, USA}

\author{W. Zhu}
\affil{National Astronomical Observatories, Chinese Academy of Science, 20A Datun Rd., Chaoyang District, Beijing 100012, China}
\affil{Max Planck Institute for Radio Astronomy, Auf dem H\"{u}gel 69, D-53121 Bonn, Germany}

\begin{abstract}
The North American Nanohertz Observatory for Gravitational Waves (NANOGrav) has observed dozens of millisecond pulsars for over a decade. We have accrued a large collection of dispersion measure (DM) measurements sensitive to the total electron content between Earth and the pulsars at each observation. All lines of sight cross through the solar wind which produces correlated DM fluctuations in all pulsars. We develop and apply techniques for extracting the imprint of the solar wind from the full collection of DM measurements in the recently released NANOGrav 11-yr data set. We filter out long time scale DM fluctuations attributable to structure in the interstellar medium and carry out a simultaneous analysis of all pulsars in our sample that can differentiate the correlated signature of the wind from signals unique to individual lines of sight. When treating the solar wind as spherically symmetric and constant in time, we find the electron number density at 1~A.U. to be $7.9\pm0.2$ cm$^{-3}$. Our data shows little evidence of long-term variation in the density of the wind. We argue that our techniques paired with a high cadence, low radio frequency observing campaign of near-ecliptic pulsars would be capable of mapping out large-scale latitudinal structure in the wind.     
\end{abstract}


\section{Introduction} \label{sec:intro}
The North American Nanohertz Observatory for Gravitational Waves (NANOGrav) has entered a second decade of precisely timing an array of millisecond pulsars (MSPs) in an effort to detect extremely low-frequency ($\sim$nHz) gravitational waves \citep{abb+18_a,abb+18_b}. As part of this effort, NANOGrav has conducted a careful accounting of the noise processes influencing our measurements, with particular attention paid to the effects of the interstellar medium (ISM) \citep{lmj+16,jml+17,lcc+15,lcc+16_1,lcc+17,wh18}. 

Pulse times of arrival (TOAs) are primarily influenced by the ISM through variable dispersive delays. At a radio frequency $\nu$, the light propagating from a pulsar to the Earth is delayed by an amount $t_d={\cal D}(t)/(K\nu^2)$, where 
\begin{equation}
{\cal D}(t)=\int_{\hat{\bf n}(t)}n_e(t,{\bf r})dl
\end{equation}
is the dispersion measure (DM), typically expressed in pc cm$^{-3}$, $K=2.41\times10^{-4}$ MHz$^{-2}$ pc cm$^{-3}$ s$^{-1}$, and $n_e(t,{\bf r})$ is the electron number density at time $t$ and position ${\bf r}$. The integration path extends from the Earth to the pulsar along the direction $\hat{\bf n}(t)$, the unit vector pointing towards the pulsar from Earth at time $t$. 

The solar wind (SW), streams of electrons flowing outward from the Sun, has a distinct and sizable influence on the DM of many pulsars. Over the course of a year, the line of sight (LOS) to a pulsar will sweep out an elliptical cone through the SW, causing annual fluctuations in DM which peak when the Sun and pulsar are in conjunction and the LOS connecting them most closely approaches the Sun. The fluctuations are larger and more peaked for pulsars closer to the ecliptic as the LOS for these pulsars more closely approaches the Sun. This is all well known. Shortly after they were discovered, pulsars were recognized as useful probes of the SW and corona \citep{cs68,h68}. The Crab Pulsar, with an ecliptic latitude $\beta=-1.24^\circ$, has been observed extensively for such applications \citep{gm69,cr72}. TEMPO and TEMPO2 \citep{nds+15,ehm06}, software packages commonly used for pulsar timing, both include constant, spherically symmetric models for the SW that attempt to account for timing perturbations it causes.  

\citet{abb+18_b} recently demonstrated that NANOGrav's sensitivity to gravitational waves has progressed to the point that our upper limit on the amplitude of the gravitational wave stochastic background depends on our choice of solar system ephemeris. To combat this undesirable model dependence, they developed tools for bridging various ephemerides, allowing the pulsar timing data itself to inform the ephemeris. 

A need has arisen for a similar treatment of the SW in which pulsar timing data can be used to inform models of the SW. As evidence for this need, \citet{agh+18} recently published important new constraints on general relativity's strong equivalence principle based on observations of a pulsar in a hierarchical triple system. They explicitly discuss issues they faced when trying to include data collected while the pulsar was close to the Sun and maximally influenced by the SW. Building on techniques used by, e.g., \citet{sns+05} and \citet{lkn+06}, \citet{agh+18} adapted the parameters of the SW to their data, but these techniques proved insufficient and systematic artifacts were left behind in their data. With observations of a lone pulsar, it is not possible to fully disentangle the influence of the ISM from that of the SW and it is difficult to constrain spatial and temporal variations in the SW. These purposes are better served by an analysis of data from a large array of pulsars, and in this work, we develop and use 
the techniques necessary to do just that. 

In nautical parlance, to ``sound'' is to measure the depth of a body of water, often in fathoms. One common sounding technique is to measure the time it takes for pulses of sound to travel from a ship, bounce off the sea floor, and return to the ship. This is not altogether dissimilar from the techniques we develop in this work: pulsed radio waves, delayed by propagation through a medium of interest---the solar wind---used to probe the distribution of that medium.

In Section~2, we describe the data we use for our analysis. In Section~3, we describe DM fluctuations caused by the SW and the ISM and the models for those fluctuations we apply to our data. In Section 4, we lay out the procedure by which we apply our DM fluctuation model to the data. In Section 5, we summarize and discuss the results of our modeling effort. Finally, in Section 6 we discuss future prospects for investigations such as this and offer some concluding remarks.


\section{Data}\label{sec:data}
To correct delays in TOAs caused by variations in DM, NANOGrav conducts at least two observations in widely separated radio frequency bands for every observing epoch of every pulsar in our array. For data collected with the 305-m William E. Gordon Telescope of the Arecibo Observatory, observations are normally taken using two separate receivers on the same day; depending on the pulsar, these are either the 430 MHz and 1.4 GHz receivers or the 1.4 GHz and 2.3 GHz receivers (or, in one case, the 327, 430, and 1400 MHz receivers). For data collected with the 100-m Robert C. Byrd Green Bank Telescope, observations are typically taken at 800 MHz and 1.4 GHz, usually within the same week. While using non-simultaneous timing observations to infer DM can induce measurement biases as the true DM varies by some amount between the measurements \citep{lcc+15,nhw+17}, these errors are typically small and we ignore them\footnote[1]{\citet{jml+17} showed that most DM variation timescales are greater than the observation cadence. Additionally, sufficiently non-simultaneous observations that occurred when the underlying DM was varying quickly because of a pulsar's proximity to the Sun have already been removed from the NANOGrav 11-yr data release to avoid large DM measurement biases.}. Furthermore, \cite{css16} recently expounded on how scattering causes light from multiple paths to converge on the observer, making the observed DM an effective average over many paths. Since scattering varies with radio frequency, so too will DM. We acknowledge but ignore this phenomenon in this work as it is particularly impactful at radio frequencies lower than those we deal with. All data are available online\footnote[2]{data.nanograv.org} and described in more detail in \citet{abb+18_a}.

A nominal DM, ${\cal D}_0$, along with a time series of perturbations to the nominal DM, $\delta{\cal D}(t_i)$ (where $t_i$ is the centroid of a bin of multifrequency TOAs), are included as free parameters in a pulsar timing model and constrained by the TOAs using generalized least-squares fitting techniques common in the practice of pulsar timing \citep{ehm06, vl13,vv14}. The linearized timing model ${\bf M}$, paired with a noise model ${\bf C}$ yield a parameter covariance matrix ${\bf C}_p=({\bf M}^T{\bf C}^{-1}{\bf M})^{-1}$. A sub-block of ${\bf C}_p$ that is symmetric about the diagonal describes covariances in the measured time series $\delta{\cal D}(t_i)$; we call this sub-block ${\bf \Sigma}$ and its $i$th diagonal element $\sigma_i^2$. Since the time series $\delta{\cal D}(t_i)$ is defined as variation about ${\cal D}_0$, it is constrained to have zero weighted mean. As such, ${\bf \Sigma}$ has a null eigenvalue and is not invertible. We define ${\bf \Xi}={\bf E}^T{\bf F}{\bf E}$, where ${\bf F}$ is a diagonal matrix containing the non-zero eigenvalues of ${\bf \Sigma}$ and the columns of ${\bf E}$ are the associated eigenvectors. Then ${\bf \Xi}$ is invertible: ${\bf \Xi}^{-1}={\bf E}^T{\bf F}^{-1}{\bf E}$.


\section{Physical Model of Dispersion Measure Fluctuations}\label{sec:model}
We treat the DM time series for a pulsar, ${\cal D}(t_i)={\cal D}_0+\delta{\cal D}(t_i)$, as a sum of two terms: contributions from the ISM, ${\cal I}(t_i)$, and contributions from the SW, ${\cal W}(t_i)$. We now discuss each of these contributions in detail, specifically how we model them.

\subsection{The Solar Wind}\label{sec:sw}
The number density of electrons in the SW can be modeled as
\begin{equation}
n_\odot(t,r,\lambda,\beta)=\sqrt{4\pi}\sum_{l,m}n_{lm}(t)Y_{lm}(\lambda,\beta)\left(\frac{1~{\rm A.U.}}{r}\right)^2,
\end{equation}
where $n_{lm}$ is a coefficient for the real spherical harmonic $Y_{lm}$. The coordinates $\lambda$ and $\beta$ are ecliptic longitude and latitude, respectively. This is a completely general form for a time-dependent field with an inverse square radial profile. An inverse square radial density profile follows from a wind ejected radially outward at a fixed velocity. \cite{imm+98}, using \emph{in situ} measurements from the \emph{Ulysses} space probe, found this inverse square scaling to hold almost exactly in far southern ecliptic latitudes. For northern latitudes, they find a steeper radial scaling for the electron density: $n_e\propto r^{-2.36}$ between approximately 1.5 and 2 A.U. Just a few solar radii ($R_\odot$) from the Sun, a variety of techniques have revealed additional contributions to $n_e$ that fall off very quickly with distance, scaling as $r^{-4}$, $r^{-6}$, or more steeply \citep{ldb98}. These additional power-law components to the radial electron density profile will only prove important for describing observations where the LOS comes within just a few degrees of the Sun. NANOGrav observations are never intentionally taken while a pulsar is particularly close to the Sun, so only a small fraction of our data could be coincidentally influenced by these regions of the SW. As such, we do not model regions of non-inverse-square scaling in this work.    

The SW is known to have strong latitudinal variation \citep{imm+97}. Near the solar activity minimum, for $|\beta|\lesssim 20^\circ$, the wind is relatively slow, producing a high electron density; the wind speed can vary a great deal with small variations of $\beta$ in this equatorial region. Further poleward, the wind is faster, producing lower electron densities, and is more-nearly constant in $\beta$. \citet{yhc+07} developed and built into TEMPO2 a DM model for the purposes of pulsar timing that incorporated both the fast and slow winds. There is some evidence for asymmetry between the northern and southern poleward winds based on \emph{in situ} measurements taken with the \emph{Ulysses} space probe, but the apparent asymmetry may be caused by evolution of the wind during the year the probe took traveling from one pole to the other \citep{imh04,ilm+08}. 

\begin{figure}
\begin{center}
\includegraphics[trim=1cm 0 0 .5cm,scale=.45]{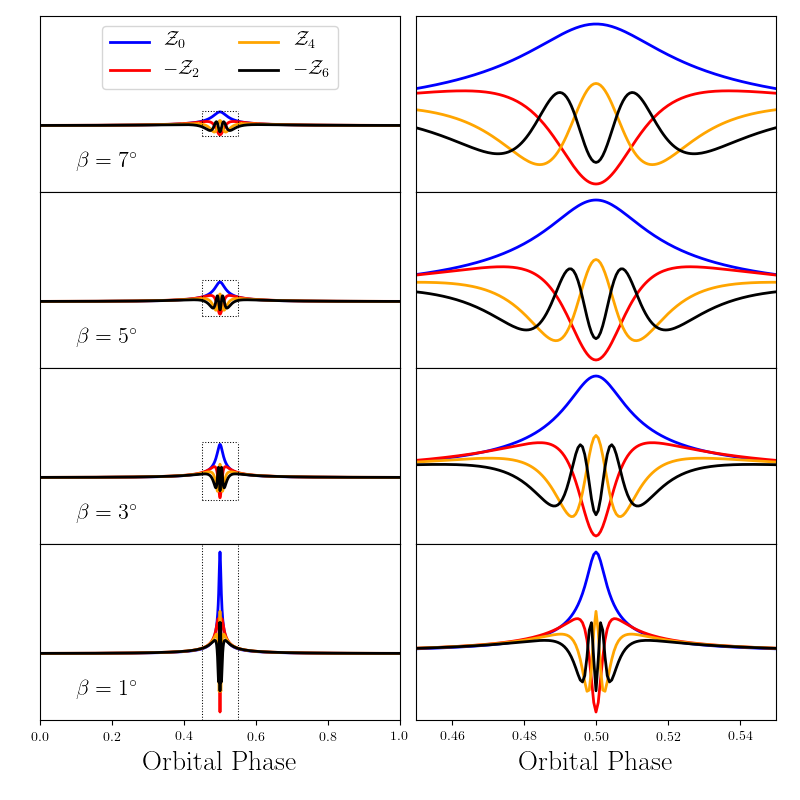}
\caption{\label{fig:Zl}Time evolution of DM fluctuations caused by solar electron clouds shaped like $m=0$ spherical harmonics of different degree $l$ for pulsars at a variety of low ecliptic latitudes. The four curves in each panel correspond to different values of $l$. We have alternated the signs of the basis functions, i.e. we show positive ${\cal Z}_0$ and negative ${\cal Z}_2$, etc., because the sign of spherical harmonics at the equator alternates as such. The amplitude of the signal decreases quickly with increasing $\beta$ (increasing as one moves upward through the panels). The panels on the left are all the same scale and show the entirety of one orbit. The panels on the right are magnifications of the black dotted boxes on the left.}
\end{center}
\end{figure}

To capture all of the known latitudinal structure in the wind, we would have to include large values of $l$ in our model. We opt to not do so and consider only models with $l=0$, and consequently, $m=0$. To demonstrate why we do this, define
\begin{equation}
\label{Zl}
{\cal Z}_l(t,\lambda,\beta)=\sqrt{4\pi}\int_{\hat{\bf n}(t)}Y_{l0}(\lambda,\beta)\left(\frac{1~{\rm A.U.}}{r}\right)^2dl.
\end{equation}
These functions encode the temporal structure of DM fluctuations attributable to moments of the SW's shape of various degrees $l$. We show several examples of ${\cal Z}_l$ in Figure \ref{fig:Zl}. There is a narrow window of orbital phase when the Sun and pulsar are near conjunction that moments with different $l$ can be differentiated from one another, and the amplitude of the signal falls off quickly as $|\beta|$ increases. Our data are not ideally suited for differentiating moments of different $l$ for two reasons: we do not sample the narrow window of orbital phase near conjunction densely enough given our approximately monthly observing cadence and only a small number of pulsars in our array are particularly close to the ecliptic.

\citet{imh04} show that the primary change in the SW over the solar cycle is that the dense wind, constrained to $|\beta|\lesssim20^\circ$ near the minimum of the solar activity cycle, spreads poleward, to as high as $|\beta|\approx 70^\circ$ during the solar activity maximum. To encapsulate this sort of time evolution in our model, we would have to incorporate moments with $l>0$. We do not consider $l>0$ for the reasons stated above, but we nonetheless consider variability in time in some of our modeling.

After these simplifications, our minimal model for the electron density in the SW is     
\begin{equation}
n_\odot(t,r,\lambda,\beta)=n_\odot(r)=n_0(t)\left(\frac{1~{\rm A.U.}}{r}\right)^2,
\end{equation}
where $n_0(t)$ is the electron number density at 1 A.U. The component of DM that can be attributed to electrons in the solar electron cloud is\footnote[3]{We assume that the solar wind is cold, diffuse, and unmagnetized. Electron number densities in the wind and the magnitude of the Sun's magnetic field make the plasma and cyclotron frequencies much lower than typical observing frequencies. Electron velocities in the cloud are sufficiently non-relativistic that the wind can be treated as cold.} ${\cal W}(t_i)=n_0(t){\cal Z}_0(t_i,\lambda,\beta)$. For all pulsars, ${\cal Z}_0$ depends exclusively on the geometry of the Earth-Sun-pulsar system, which is precisely measured through timing measurements. 

In the frequency domain, fluctuations in DM from the SW appear with a fundamental frequency of 1 yr$^{-1}$. The phase of the signature is known. As the ecliptic latitude of a pulsar approaches zero, the duty cycle of the periodic SW signature becomes smaller, transferring power from the fundamental frequency into higher harmonics. In the coming discussion, we demonstrate that the ISM can produce similar periodic fluctuations with high harmonic content, but that with an entire array of pulsars, the signatures can be readily disentangled.

\subsection{The Interstellar Medium}\label{sec:ism}
 
The ISM is an inhomogeneous medium with ionized density structures following an approximately Kolomogorov scaling law, i.e. the power spectrum of spatial wave numbers, $q$, is proportional to $q^{-\kappa}$ with $\kappa\approx11/3$. Embedded in this turbulent Kolomogov medium, there are discrete density structures such as magnetically collimated filaments and plasma lenses that can cause DM variation events and additional chromatic timing behavior inconsistent with the expectations of a Kolomogorov medium \citep[e.g.][]{cks+15,leg+18}. We do not incorporate such structures into our modeling because these discrete structures appear only rarely.    

To model the influence of the turbulent ISM on observations of pulsars, material in the ISM is commonly described as being confined to a thin screen between the Earth and pulsar, transverse to the LOS. Using the techniques of \citet{css16}, we have simulated the electromagnetic phase perturbation, $\phi$, generated by propagation of light through thin screens of a Kolomogorov medium; phase perturbations $\phi(t)$ are related to DM perturbations as ${\cal D}(t)=-\nu\phi(t)/(c r_e)$, where $c$ is the speed of light and $r_e$ is the classical electron radius \citep{r90}.    

The trajectory that the LOS cuts through a screen depends on the distance of the screen and pulsar from the solar system barycenter (SSB), $D_s$ and $D_p$ respectively. The ecliptic coordinates and proper motion of the pulsar also affect the trajectory. For simplicity, we will assume that the screen is at rest relative to the SSB. Further, assume the Earth's orbit, ${\bf x}_e(t)$, is circular and perfectly confined to the ecliptic, i.e.
\begin{equation}
{\bf x}_e(t)=D_1\left[\begin{array}{c}
\cos{[\omega_1(t-t_R)]}\\
\sin{[\omega_1(t-t_R)]}\\
0
\end{array}\right]
\end{equation}
where $\omega_1=2\pi$ yr$^{-1}$, $D_1=1$ A.U., and $t_R$ is a reference epoch, specifically an autumnal equinox. 

At some reference epoch, $t_0$, the position of a pulsar relative to the SSB is
\begin{equation}
{\bf x}_p(t_0)=D_p\left[\begin{array}{c}
\cos{\lambda_0}\cos{\beta_0}\\
\sin{\lambda_0}\cos{\beta_0}\\
\sin{\beta_0}
\end{array}\right],
\end{equation}
where $\lambda_0$ and $\beta_0$ are the pulsar's ecliptic latitude and longitude at $t_0$. Since accelerations are small, we will ignore them and take the rates of change of the pulsar's ecliptic coordinates as $\dot{\beta}=\mu_\beta$ and $\dot{\lambda}=\mu_\lambda/\cos{\beta}$ where $\mu_\beta$ and $\mu_\lambda$ are the components of the pulsar's proper motion in ecliptic coordinates.

\begin{figure*}
\begin{center}
\includegraphics[scale=.9]{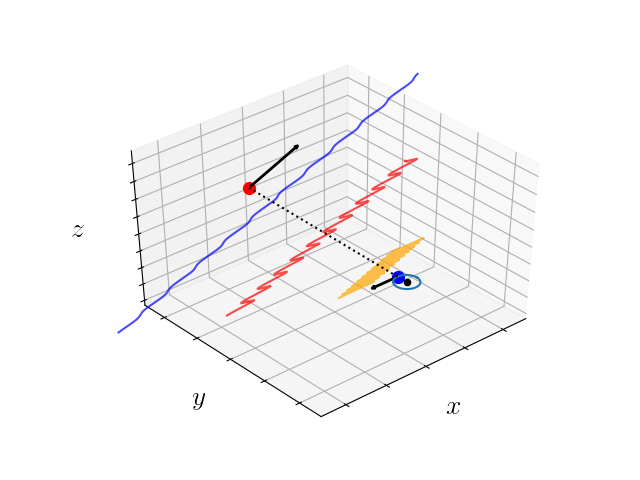}
\includegraphics[scale=.58]{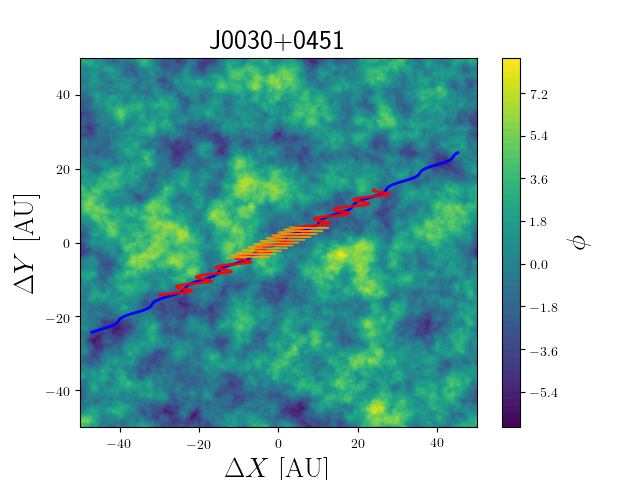}
\includegraphics[scale=.52]{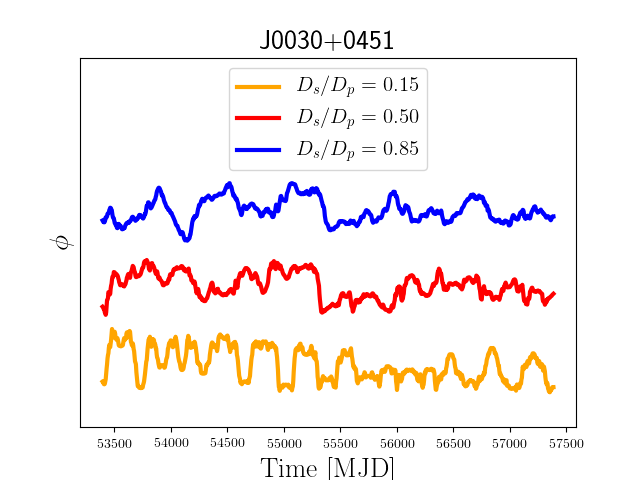}
\caption{\label{fig:screenInSpace}{\bf Top:} The Earth (blue dot) orbits the solar system barycenter (black dot) and the pulsar (red dot) displays proper motion transverse to the LOS. The orange, red, and blue curves represent the trajectory the LOS cuts through transverse planes at varying distance between the Earth and pulsar (not to scale). These curves were generated using the ecliptic coordinates and proper motion for J0030$+$0451. {\bf Bottom Left:} A Kolomogorov phase perturbation screen and the trajectory traced through the screen if it is placed at different distances between the Earth and pulsar. {\bf Bottom Right:} The phase perturbation from the Kolomogorov screen sampled along the three different trajectories. The nearest screen (orange) leads to quasi-periodic fluctuations in $\phi$. This quasi-periodicity is noticeably reduced for more distant screens. The vertical offset between the curves is put in by hand for visual clarity.}
\end{center}
\end{figure*}

\begin{figure}
\begin{center}
\includegraphics[scale=.55]{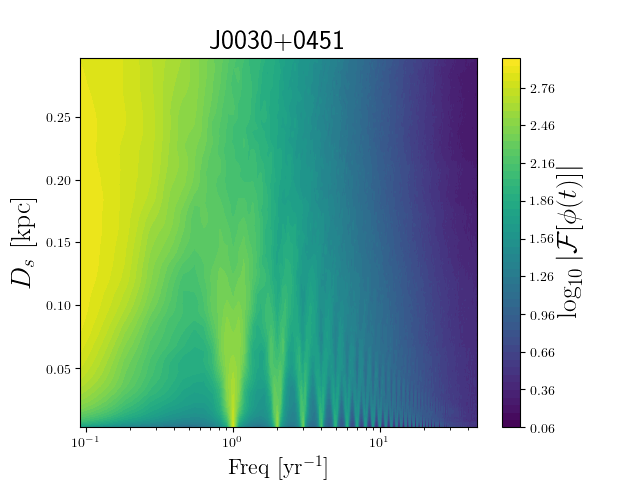}
\includegraphics[scale=.55]{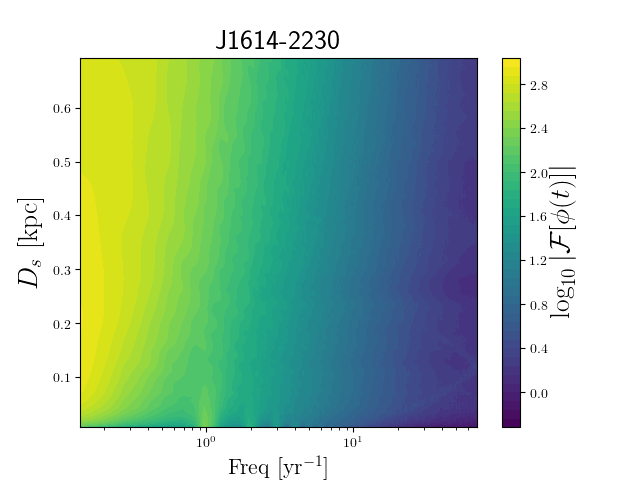}
\caption{\label{fig:spatialSpectra} The anticipated spectral power in DM fluctuations from screens at varied distances from the SSB $D_s$ for two example pulsars with low ecliptic latitudes. The position, proper motion, and distance of a pulsar strongly influence the anticipated result. J0030$+$0451, a nearby, low proper motion pulsar, can show pronounced quasi-periodic fluctuations for sufficiently close screens. Faster moving and more distant J1614$-$2230 is less prone to such quasi-periodic fluctuations. The low-frequency fluctuation power is a feature common to all pulsars; it is associated with sampling distinct regions of the screen over time.}
\end{center}
\end{figure}

Define $\hat{\bf X}$ and $\hat{\bf Y}$, orthogonal vectors spanning planes transverse to the LOS at $t_0$:
\begin{eqnarray}
\hat{\bf X}&=&\left[\begin{array}{c}
\sin{\lambda_0}\\
-\cos{\lambda_0}\\
0
\end{array}\right],\\
\hat{\bf Y}&=&\left[\begin{array}{c}
-\cos{\lambda_0}\sin{\beta_0}\\
-\sin{\lambda_0}\sin{\beta_0}\\
\cos{\beta_0}
\end{array}\right].
\end{eqnarray}
We specify positions in the screen with coordinates $\Delta X$ and $\Delta Y$ along $\hat{\bf X}$ and $\hat{\bf Y}$ respectively. If $D_s\gg D_1$, the location at which the LOS intersects the screen will be 
\begin{equation}
{\bf x}_s(t)\approx\frac{D_s}{D_p}{\bf x}_p(t)+\left(1-\frac{D_s}{D_p}\right){\bf x}_e(t).
\end{equation}
Projected onto the basis spanning the screen, 
\begin{eqnarray}
\label{eq:xTrajectory}
{\bf x}_s(t)\cdot\hat{\bf X}&\approx&-D_s\mu_\lambda(t-t_0)\nonumber\\&&-D_1\left(1-\frac{D_s}{D_p}\right)\sin{[\omega_1(t-t_R)-\lambda_0]},\nonumber\\
\end{eqnarray}
and
\begin{eqnarray}
\label{eq:yTrajectory}
{\bf x}_s(t)\cdot\hat{\bf Y}&\approx&D_s\mu_\beta(t-t_0)\nonumber\\&&-D_1\sin{\beta_0}\left(1-\frac{D_s}{D_p}\right)\cos{[\omega_1(t-t_R)-\lambda_0]}.\nonumber\\
\end{eqnarray}
In each component of the screen trajectory, there is one term that grows linearly in time and is proportional to the product of $D_s$ and a component of proper motion; the other term oscillates annually and has a larger amplitude for smaller values of $D_s$. The relative scale of these two terms influences the qualitative shape of the trajectory the LOS cuts through the screen and, consequently, the spectral properties of the DM fluctuations caused by sampling the screen along that trajectory. This is demonstrated in Figures~\ref{fig:screenInSpace} and \ref{fig:spatialSpectra}.     

In the top panel of Figure~\ref{fig:screenInSpace} we depict the trajectory the LOS tracks across screens placed at different distances from the Earth for J0030$+$0451. The bottom left panel of Figure~\ref{fig:screenInSpace} shows those three different trajectories projected onto a single realization of a Kolomogorov phase screen. The bottom right panel shows the value of $\phi$ from the phase screen evaluated along those trajectories (color coded). The nearest screen yields fluctuations in $\phi$ that are noticeably quasi-periodic with an approximately annual fundamental periodicity. As the screen distance is increased, this quasi-periodicity begins to vanish. Figure~\ref{fig:spatialSpectra} maps out the influence of $D_s$ on the spectral properties of fluctuations in $\phi$. We vary $D_s$ between 1\% and 99\% of $D_p$, sample $\phi$ along the trajectory the LOS cuts through the screen, and compute the modulus of the Fourier transform of the resultant time series, $|{\cal F}(\phi)|$. The surfaces in Figure \ref{fig:spatialSpectra} show the results of this calculation averaged over 100 screen realizations for PSRs J0030$+$0451 and J1614$-$2230.

The results in Figure \ref{fig:spatialSpectra} are qualitatively different for the two pulsars. The magnitude of proper motion for J0030$+$0451 is approximately 6 mas yr$^{-1}$, substantially smaller than the proper motion of J1614$-$2230 which is approximately 32 mas yr$^{-1}$. Additionally, the distance to J0030$+$0451 is approximately 0.3 kpc as opposed to 0.65 kpc for J1614$-$2230 \citep{mnf+16}. Since J0030$+$0451 is closer than J1614$-$2230 and has less proper motion, the linearly growing terms in Equations~\ref{eq:xTrajectory} and \ref{eq:yTrajectory} are typically less significant for J0030$+$0451, making the annual terms more important. These statements about the shape of the trajectories the LOS cuts through screens, when mapped to the spectral properties of temporal variations in $\phi$, mean that one expects J0030$+$0451 to show more quasi-periodic variation from the ISM than J1614$-$2230. For $D_s\lesssim$ 0.1 kpc, J0030$+$0451 displays strong fluctuations at approximately 1 yr$^{-1}$ with additional power at numerous higher harmonics. J1614$-$2230 is relatively free from such quasiperiodicity. The low-frequency ``red'' power (more power at lower frequencies) visible for all but the lowest values of $D_s$ is associated with the linear terms in Equations~\ref{eq:xTrajectory} and \ref{eq:yTrajectory}, stochastic fluctuations caused by the sampling region drifting across a Kolomogorov screen. 


\section{Implementation}\label{sec:implementation} 

We describe our DM time series as a superposition of SW and ISM contributions, ${\cal W}$ and ${\cal I}$, respectively:
\begin{equation}
\label{model}
{\cal D}_0+\delta{\cal D}={\cal W}+{\cal I}.
\end{equation}
Based on our analysis in Section \ref{sec:ism}, we further subdivide ${\cal I}$ into the sum of two terms: ${\cal R}$, a stochastic red process confined to frequencies below 1 yr$^{-1}$, and ${\cal P}$, a periodic sinusoid of unspecified amplitude and phase with a frequency of 1 yr$^{-1}$. This is, of course, an approximate description. The low-frequency red power extends to frequencies above 1 yr$^{-1}$, but specifically because it is red, power in those higher frequencies will be sub-dominant. Power in the fundamental harmonic of the quasi-periodic oscillations can be at frequencies near, but not at, 1 yr$^{-1}$. Additionally, harmonics of 1 yr$^{-1}$ may be present in the periodic signals associated with the ISM. We ignore these higher harmonics in our description of ${\cal I}$ because they contain less power than the fundamental and they only matter if there are very nearby screens between Earth and pulsars with low proper motion.

Our aim in this work is to make inferences about the SW with our DM measurements. From this perspective, the SW signal is contaminated by the ISM signal, part of which, ${\cal R}$, is stochastic and red. To mitigate ${\cal R}$, we perform a weighted Gaussian convolution to separate our DM time series $\delta{\cal D}$ into low-frequency and high-frequency contributions, $\delta\bar{\cal D}$ and $\delta\widetilde{\cal D}$, respectively. In detail,
\begin{equation}
\label{eq:smoothing}
\delta\bar{\cal D}(t_i)=\left[\sum_jw_{ij}\right]^{-1}\sum_jw_{ij}\delta{\cal D}(t_j),
\end{equation}
where,
\begin{equation}
\label{eq:weights}
w_{ij}=\frac{1}{\sigma_j^2}\exp{\left[-\frac{(t_i-t_j)^2}{2\tau^2}\right]},
\end{equation}
and $\tau$ is a smoothing time scale. This defines a linear operator ${\bf L}$ such that $\delta\bar{\cal D}={\bf L}\delta{\cal D}$. Also define ${\bf H}={\bf I}-{\bf L}$ where ${\bf I}$ is the identity. Then $\delta\widetilde{\cal D}={\bf H}\delta{\cal D}$. Note that the nominal DM, ${\cal D}_0$, is a constant signal, so ${\bf L}{\cal D}_0={\cal D}_0$ and ${\bf H}{\cal D}_0=0$. 

We have fixed the time scale $\tau=\sqrt{2\log{2}}/\pi$ yrs; convolution of a time series with a Gaussian of width $\tau$ is equivalent to multiplying the Fourier transform of that time series by a Gaussian centered at zero with a half width at half max of 0.5 yr$^{-1}$. Power at frequencies above 0.5 yr$^{-1}$ is strongly attenuated, making $\delta\bar{\cal D}$ a smoothed version of $\delta{\cal D}$ largely devoid of periodic signals with frequencies at or above 1 yr$^{-1}$. 

We choose the above value of $\tau$ so that ${\bf H}{\cal R}\approx 0$. In practice, we write ${\cal P}=a{\cal S}+b{\cal C}$, a linear combination of an annual sinusoid, ${\cal S}=\sin{[\omega_1(t-t_R)]}$, and cosinusoid, ${\cal C}=\cos{[\omega_1(t-t_R)]}$, of unspecified amplitudes $a$ and $b$. This can alternatively be parameterized as ${\cal P}=A\cos{[\omega_1(t-t_R)-\Phi]}$, where $A^2=a^2+b^2$ and $\Phi=\arctan2{(a,b)}$. Multiplying Equation~\ref{model} through by ${\bf H}$ yields
\begin{eqnarray}
\label{eq:filteredData}
\delta\widetilde{\cal D}&=&{\bf H}({\cal W}+{\cal P}+{\cal R}),\nonumber\\
&\approx&n_0(t)\widetilde{\cal Z}_0+a\widetilde{\cal S}+b\widetilde{\cal C},
\end{eqnarray}
where $\widetilde{\cal Z}_0={\bf H}{\cal Z}_0$ and $\widetilde{\cal S}$ and $\widetilde{\cal C}$ are similarly defined. As a reminder, ${\cal Z}_0$ is defined by Eq. \ref{Zl}. Though ${\cal Z}_0$, ${\cal S}$, and ${\cal C}$ primarily consist of power at or above frequencies of 1 yr$^{-1}$, they are somewhat modified by the high-pass filter ${\bf H}$. To quantify that modification, define $\Delta{\cal Z}_0={\cal Z}_0-\widetilde{\cal Z}_0$. Similarly define $\Delta{\cal S}$ and $\Delta{\cal C}$. These corrections are useful because they connect idealized basis elements like ${\cal Z}_0$ to their filtered counterparts ($\widetilde{\cal Z}_0$ in this case) which depend on the data.

Consider the simple case where $n_0$ is constant. In this case, we model the ${\bf H}$-filtered DM time series for individual pulsars as a linear combination of three basis elements. We constrain the coefficients $n_0$, $a$, and $b$ with generalized least-squares techniques. Define a so-called ``design matrix'' ${\bf M}=[\widetilde{\cal Z}_0,\widetilde{\cal S},\widetilde{\cal C}]$. Then the best-fit values for the coefficients are
\begin{equation}
\left[\begin{array}{c}
\hat{n}_0\\\hat{a}\\\hat{b}
\end{array}\right]=\left({\bf M}^T{\bf \Xi}^{-1}{\bf M}\right)^{-1}{\bf M}^T{\bf \Xi}^{-1}\delta\widetilde{\cal D},
\end{equation}
where ${\bf \Xi}^{-1}$ was defined in Section~\ref{sec:data}. Least-squares analysis like this has been done in this context before. \citet{sns+05}, in a study of PSR J1713+0747, found $\hat{n}_0=5\pm4$~cm$^{-3}$. In a similar study of PSR J0030+0451, a pulsar much nearer the ecliptic than J1713+0747, \citet{lkn+06} found $\hat{n}_0=6.9\pm2.1$~cm$^{-3}$. Both authors assumed a constant value for $n_0$, so we will consider that case as well for comparison, but neither attempted to mitigate the stochastic low-frequency or periodic signatures potentially produced by the ISM.

\begin{deluxetable*}{l|ccc|ccc|ccc|ccc}
\tablewidth{0pc}
\tablecaption{Amplitude and Phase of Annual Sinusoidal DM Fluctuations from the ISM}
\tablehead{
	\colhead{PSR}&
	\multicolumn{3}{c}{$\hat{a}\times10^5$}&
    \multicolumn{3}{c}{$\hat{b}\times10^5$}&
    \multicolumn{3}{c}{$\hat{A}\times10^5$}&
    \multicolumn{3}{c}{$\hat{\Phi}$}\\
    \colhead{}&
    \multicolumn{3}{c}{[pc cm$^{-3}$]}&
    \multicolumn{3}{c}{[pc cm$^{-3}$]}&
    \multicolumn{3}{c}{[pc cm$^{-3}$]}&
    \multicolumn{3}{c}{[deg]}
	}
\startdata
J0023$+$0923    &       21.2    &$\pm$& 7.2     &       $-$9.1  &$\pm$& 6.3     &       23.1    &$\pm$& 7.2     &       113.1   &$\pm$& 16.0    \\
J0030$+$0451    &       $-$5.2  &$\pm$& 3.4     &       $-$1.7  &$\pm$& 3.8     &       5.5     &$\pm$& 3.7     &       $-$108.1        &$\pm$& 37.2    \\
J0340$+$4130    &       6.4     &$\pm$& 5.0     &       $-$12.7 &$\pm$& 6.0     &       14.3    &$\pm$& 5.5     &       153.0   &$\pm$& 22.3    \\
J0613$-$0200    &       2.3     &$\pm$& 1.6     &       10.6    &$\pm$& 1.7     &       10.9    &$\pm$& 1.7     &       12.6    &$\pm$& 8.4     \\
J0636$+$5128    &       $-$6.1  &$\pm$& 4.0     &       12.7    &$\pm$& 5.7     &       14.1    &$\pm$& 5.3     &       $-$25.9 &$\pm$& 18.1    \\
J0645$+$5158    &       6.4     &$\pm$& 3.4     &       2.5     &$\pm$& 2.4     &       6.8     &$\pm$& 3.1     &       68.4    &$\pm$& 23.2    \\
J0740$+$6620    &       0.0     &$\pm$& 10.0    &       $-$9.6  &$\pm$& 11.1    &       9.6     &$\pm$& 11.1    &       179.9   &$\pm$& 59.5    \\
J0931$-$1902    &       6.1     &$\pm$& 8.7     &       $-$30.7 &$\pm$& 8.3     &       31.3    &$\pm$& 8.3     &       168.6   &$\pm$& 16.1    \\
J1012$+$5307    &       6.6     &$\pm$& 2.8     &       2.8     &$\pm$& 3.8     &       7.2     &$\pm$& 3.3     &       67.1    &$\pm$& 27.0    \\
J1024$-$0719    &       5.6     &$\pm$& 3.6     &       0.4     &$\pm$& 2.4     &       5.6     &$\pm$& 3.5     &       85.6    &$\pm$& 26.4    \\
J1125$+$7819    &       158.8   &$\pm$& 96.4    &       $-$91.3 &$\pm$& 92.8    &       183.1   &$\pm$& 84.6    &       119.8   &$\pm$& 32.4    \\
J1453$+$1902    &       $-$13.7 &$\pm$& 18.1    &       24.3    &$\pm$& 22.4    &       27.9    &$\pm$& 22.5    &       $-$29.3 &$\pm$& 36.6    \\
J1455$-$3330    &       16.7    &$\pm$& 9.1     &       $-$6.2  &$\pm$& 9.0     &       17.8    &$\pm$& 9.2     &       110.3   &$\pm$& 28.5    \\
J1600$-$3053    &       $-$5.0  &$\pm$& 2.7     &       0.3     &$\pm$& 2.8     &       5.0     &$\pm$& 2.7     &       $-$85.8 &$\pm$& 32.8    \\
J1614$-$2230    &       $-$3.6  &$\pm$& 2.8     &       11.2    &$\pm$& 2.6     &       11.8    &$\pm$& 2.7     &       $-$17.9 &$\pm$& 13.4    \\
J1640$+$2224    &       $-$3.5  &$\pm$& 1.0     &       $-$3.7  &$\pm$& 1.1     &       5.1     &$\pm$& 1.2     &       $-$136.2        &$\pm$& 11.5    \\
J1643$-$1224    &       $-$29.0 &$\pm$& 11.4    &       28.5    &$\pm$& 11.9    &       40.7    &$\pm$& 11.0    &       $-$45.4 &$\pm$& 17.2    \\
J1713$+$0747    &       $-$1.5  &$\pm$& 0.6     &       0.3     &$\pm$& 0.7     &       1.6     &$\pm$& 0.7     &       $-$78.7 &$\pm$& 26.0    \\
J1738$+$0333    &       $-$28.4 &$\pm$& 13.0    &       22.4    &$\pm$& 11.3    &       36.2    &$\pm$& 12.1    &       $-$51.7 &$\pm$& 19.4    \\
J1741$+$1351    &       3.7     &$\pm$& 3.4     &       9.3     &$\pm$& 3.1     &       10.1    &$\pm$& 3.7     &       22.0    &$\pm$& 16.3    \\
J1744$-$1134    &       $-$3.9  &$\pm$& 2.7     &       $-$1.5  &$\pm$& 2.9     &       4.2     &$\pm$& 2.8     &       $-$111.2        &$\pm$& 38.5    \\
J1747$-$4036    &       29.4    &$\pm$& 39.1    &       68.1    &$\pm$& 35.1    &       74.2    &$\pm$& 36.2    &       23.3    &$\pm$& 29.4    \\
J1832$-$0836    &       $-$7.6  &$\pm$& 6.8     &       $-$44.3 &$\pm$& 6.2     &       45.0    &$\pm$& 6.3     &       $-$170.2        &$\pm$& 8.5     \\
J1853$+$1303    &       36.8    &$\pm$& 28.6    &       $-$1.2  &$\pm$& 23.2    &       36.8    &$\pm$& 28.8    &       91.9    &$\pm$& 35.9    \\
B1855$+$09      &       $-$2.6  &$\pm$& 2.2     &       1.8     &$\pm$& 2.3     &       3.2     &$\pm$& 2.2     &       $-$54.7 &$\pm$& 41.3    \\
J1903$+$0327    &       $-$110.3        &$\pm$& 27.5    &       60.3    &$\pm$& 32.7    &       125.7   &$\pm$& 25.1    &       $-$61.3 &$\pm$& 15.7    \\
J1909$-$3744    &       $-$3.3  &$\pm$& 0.7     &       5.9     &$\pm$& 0.9     &       6.8     &$\pm$& 0.9     &       $-$29.4 &$\pm$& 6.8     \\
J1910$+$1256    &       $-$1.4  &$\pm$& 11.1    &       28.8    &$\pm$& 9.6     &       28.8    &$\pm$& 9.6     &       $-$2.9  &$\pm$& 22.0    \\
J1911$+$1347    &       $-$11.3 &$\pm$& 2.5     &       16.5    &$\pm$& 2.1     &       20.0    &$\pm$& 2.3     &       $-$34.4 &$\pm$& 6.7     \\
J1918$-$0642    &       $-$0.2  &$\pm$& 1.9     &       7.0     &$\pm$& 2.2     &       7.0     &$\pm$& 2.1     &       $-$2.2  &$\pm$& 15.5    \\
J1923$+$2515    &       $-$2.2  &$\pm$& 5.2     &       $-$12.3 &$\pm$& 4.8     &       12.5    &$\pm$& 5.2     &       $-$169.7        &$\pm$& 21.9    \\
B1937$+$21      &       $-$20.5 &$\pm$& 4.7     &       13.9    &$\pm$& 4.7     &       24.8    &$\pm$& 4.6     &       $-$55.8 &$\pm$& 11.2    \\
J1944$+$0907    &       $-$7.3  &$\pm$& 6.6     &       26.5    &$\pm$& 7.1     &       27.5    &$\pm$& 6.4     &       $-$15.4 &$\pm$& 15.1    \\
B1953$+$29      &       $-$2.4  &$\pm$& 19.5    &       21.9    &$\pm$& 20.1    &       22.1    &$\pm$& 19.2    &       $-$6.4  &$\pm$& 53.1    \\
J2010$-$1323    &       $-$2.7  &$\pm$& 2.5     &       1.8     &$\pm$& 2.9     &       3.2     &$\pm$& 2.4     &       $-$55.5 &$\pm$& 52.1    \\
J2017$+$0603    &       $-$0.2  &$\pm$& 5.0     &       $-$41.4 &$\pm$& 6.3     &       41.4    &$\pm$& 6.3     &       $-$179.6        &$\pm$& 6.9     \\
J2033$+$1734    &       8.1     &$\pm$& 13.6    &       21.4    &$\pm$& 17.9    &       22.9    &$\pm$& 19.0    &       20.7    &$\pm$& 30.3    \\
J2043$+$1711    &       2.2     &$\pm$& 3.6     &       $-$8.3  &$\pm$& 3.1     &       8.6     &$\pm$& 3.1     &       165.1   &$\pm$& 24.2    \\
J2145$-$0750    &       1.0     &$\pm$& 8.0     &       5.7     &$\pm$& 7.5     &       5.8     &$\pm$& 7.6     &       9.9     &$\pm$& 77.8    \\
J2214$+$3000    &       $-$81.7 &$\pm$& 29.2    &       62.6    &$\pm$& 27.7    &       102.9   &$\pm$& 28.9    &       $-$52.5 &$\pm$& 15.6    \\
J2229$+$2643    &       $-$2.8  &$\pm$& 3.5     &       3.8     &$\pm$& 4.1     &       4.7     &$\pm$& 3.7     &       $-$36.7 &$\pm$& 48.6    \\
J2234$+$0611    &       $-$6.6  &$\pm$& 4.8     &       $-$2.7  &$\pm$& 6.0     &       7.1     &$\pm$& 5.6     &       $-$112.4        &$\pm$& 42.2    \\
J2234$+$0944    &       13.0    &$\pm$& 19.8    &       62.2    &$\pm$& 21.8    &       63.5    &$\pm$& 20.6    &       11.8    &$\pm$& 18.9    \\
J2302$+$4442    &       9.2     &$\pm$& 16.3    &       19.5    &$\pm$& 14.1    &       21.5    &$\pm$& 15.0    &       25.2    &$\pm$& 41.4    \\
J2317$+$1439    &       0.9     &$\pm$& 2.0     &       8.0     &$\pm$& 2.3     &       8.1     &$\pm$& 2.3     &       6.9     &$\pm$& 14.1   
\enddata
\tablecomments{Best-fit amplitude and 1-$\sigma$ uncertainty on the amplitudes, $\hat{a}$ and $\hat{b}$, of sinusoidal annual fluctuations in DM associated with the ISM. The quantities $\hat{A}$ and $\hat{\Phi}$ are functions of $\hat{a}$ and $\hat{b}$ as described in the text preceding Eq. \ref{eq:filteredData}.}
\end{deluxetable*}

The NANOGrav 11-yr data set contains DM time series for $N=45$ pulsars, and we can leverage the whole data set to constrain $n_0$. To this end, we append an index to our basis elements to indicate which pulsar in the array we are referring to, i.e. $\widetilde{\cal Z}_{0,1}$, $\widetilde{\cal S}_1$, and $\widetilde{\cal C}_1$ correspond to the first pulsar in our array. Define a global design matrix
\begin{equation}
{\bf M_G}=
\left[\begin{array}{cccccccc}
\widetilde{\cal Z}_{0,1}&\widetilde{\cal S}_1&\widetilde{\cal C}_1&0&0&\cdots&0&0\\
\widetilde{\cal Z}_{0,2}&0&0&\widetilde{\cal S}_2&\widetilde{\cal C}_2&\cdots&0&0\\
\vdots&\vdots&\vdots&\vdots&\vdots&\ddots&\vdots&\vdots\\
\widetilde{\cal Z}_{0,N}&0&0&0&0&\cdots&\widetilde{\cal S}_N&\widetilde{\cal C}_N
\end{array}\right].
\end{equation}
There are $2N+1$ columns in ${\bf M_G}$ and the number of rows is the total number of DM measurements summed over all pulsars\footnote[4]{For PSR J1713$+$0747, we removed 11 DM measurements between MJDs 54710 and 55080. During this time, there was an extreme scattering event observed in this pulsar, causing DM evolution not describable within our framework \citep{leg+18}.}, $N_{\rm DM}=3321$. Similarly, define a block-diagonal global inverse covariance matrix
\begin{equation}
{\bf \Xi}_{\bf G}^{-1}=
\left[\begin{array}{ccc}
{\bf \Xi}_1^{-1}&\cdots&0\\
\vdots&\ddots&\vdots\\
0&\cdots&{\bf \Xi}_N^{-1}
\end{array}
\right].
\end{equation}
Then
\begin{equation}
\left[\begin{array}{c}
\hat{n}_0\\
\hat{a}_1\\
\hat{b}_1\\
\vdots\\
\hat{a}_N\\
\hat{b}_N
\end{array}\right]={\bf C}_{p,{\bf G}}{\bf M_G}^T{\bf \Xi}_{\bf G}^{-1}\delta\widetilde{\cal D}_{{\bf G}},
\end{equation}
where $\delta\widetilde{\cal D}_{{\bf G}}^T=[\delta\widetilde{\cal D}_{1}^T\cdots\delta\widetilde{\cal D}_{N}^T]$ and the global parameter covariance matrix ${\bf C}_{p,{\bf G}}=({\bf M}_{\bf G}^T{\bf \Xi}_{\bf G}^{-1}{\bf M}_{\bf G})^{-1}$. 

With a slight elaboration of this least-squares framework, we also test for variations in $n_0$ over time. We define a grid of $N_T=13$ times $T_i$ with one-year spacing; the latest of them is MJD 53788---one day after the final observation in the 11-yr data set---and the earliest is precisely 12 years earlier, spanning the full 11-yr data set (which actually spans approximately 11.4 yr). We treat $n_0$ as piecewise constant between grid points, meaning we allow it to take on $N_T-1=12$ different values. Rather than a single basis element ${\cal Z}_0$ being used to describe SW fluctuations, we now use $N_T-1$ basis elements ${\cal Z}_0(T_i,T_{i+1})$ that are equal to ${\cal Z}_0$ between $T_i$ and $T_{i+1}$ and zero otherwise. These are then high-pass filtered to produce $\widetilde{\cal Z}_0(T_i,T_{i+1})$. The solar basis elements are then stacked into the first $N_T-1$ columns of ${\bf M}_G$. 

\section{Results}\label{sec:results}

Our final model for the DM variations in a pulsar is
\begin{equation}
\label{eq:fullModel}
\delta{\cal D}_M=\delta\bar{\cal D}+\hat{n}_0\widetilde{\cal Z}_0+\hat{a}\widetilde{\cal S}+\hat{b}\widetilde{\cal C}.
\end{equation}
We first consider the case where $n_0$ is constant. After carrying out the analysis described above, the values of $\hat{a}$ and $\hat{b}$ for each of the 45 pulsars in the NANOGrav 11-yr data set, along with their uncertainties, are given in Table 1. As noted above, the annual DM fluctuations from the ISM, ${\cal P}=a{\cal S}+b{\cal C}$, can alternatively be parameterized as ${\cal P}=A\cos{[\omega_1(t-t_R)-\Phi]}$ with $A^2=a^2+b^2$ and $\Phi=\arctan2{(a,b)}$. We also give best fit values of these parameters, $\hat{A}$ and $\hat{\Phi}$, in Table 1. The ability to differentiate the annual DM fluctuations caused by structure in the ISM from those caused by the SW is greatly facilitated by the techniques developed in this work, relying on the many lines of sight made accessible by a full pulsar timing array.  

Our best-fit value for the electron density in the SW at 1 A.U. when it is assumed to be constant is $\hat{n}_0=7.9\pm0.2$~cm$^{-3}$. The default value of $n_0$ used by TEMPO2 is 4 cm$^{-3}$; the default value used by TEMPO is 10~cm$^{-3}$ \citep{ehm06,nds+15}. Our result indicates that the default model for the SW in TEMPO2 will underestimate dispersive delays while TEMPO will overestimate them. 

We can compare our result for $\hat{n}_0$ to a long line of pulsar-based inferences about the SW. \citet{gm69} observed the Crab Pulsar with the old 300-foot telescope of the National Radio Astronomy Observatory in Green Bank during an occultation by the solar corona and failed to successfully detect variations in DM from the SW at all. With the more sensitive Arecibo telescope and using timing techniques more closely resembling those used today, \citet{cr72} observed the Crab Pulsar through occultation in 1969 and 1970. They not only detected DM variations from the SW\footnote[5]{\citet{cr72} measured the electron density of the SW at 10 $R_\odot$ to be 7000$\pm$600 cm$^{-3}$. Given the non-inverse-square scaling of the electron density they found within 20 $R_\odot$ and that is known to exist from other studies, it is not straightforward to compare their measured electron density to ours which we have referenced to 1 A.U.}, they found that between 5 and 20 solar radii, the SW was not yet free-streaming, and the electron density scaled as the radius to the $-2.9\pm0.2$ power. As we have discussed, in a study of PSR J1713+0747, \citet{sns+05} found $n_0=5\pm4$~cm$^{-3}$---a marginal detection. In a study of PSR J0030+0451, with an ecliptic latitude of approximately 1.5$^\circ$ (the ecliptic latitude of PSR J1713+0747 is approximately 30$^\circ$), \citet{lkn+06} measured $n_0=6.9\pm2.1$ cm$^{-3}$. The results from \citet{sns+05} and \citet{lkn+06} are consistent with our result; our much-increased precision is attributable to improvements in hardware at Arecibo and GBT \citep{drd+08,rdf+09} and to the techniques we have developed here to combine measurements from many pulsars. Furthermore, our techniques allow us to mitigate potential bias in SW measurements caused by annual fluctuations in DM from the ISM.

\begin{figure*}
\begin{center}
\includegraphics[scale=.5]{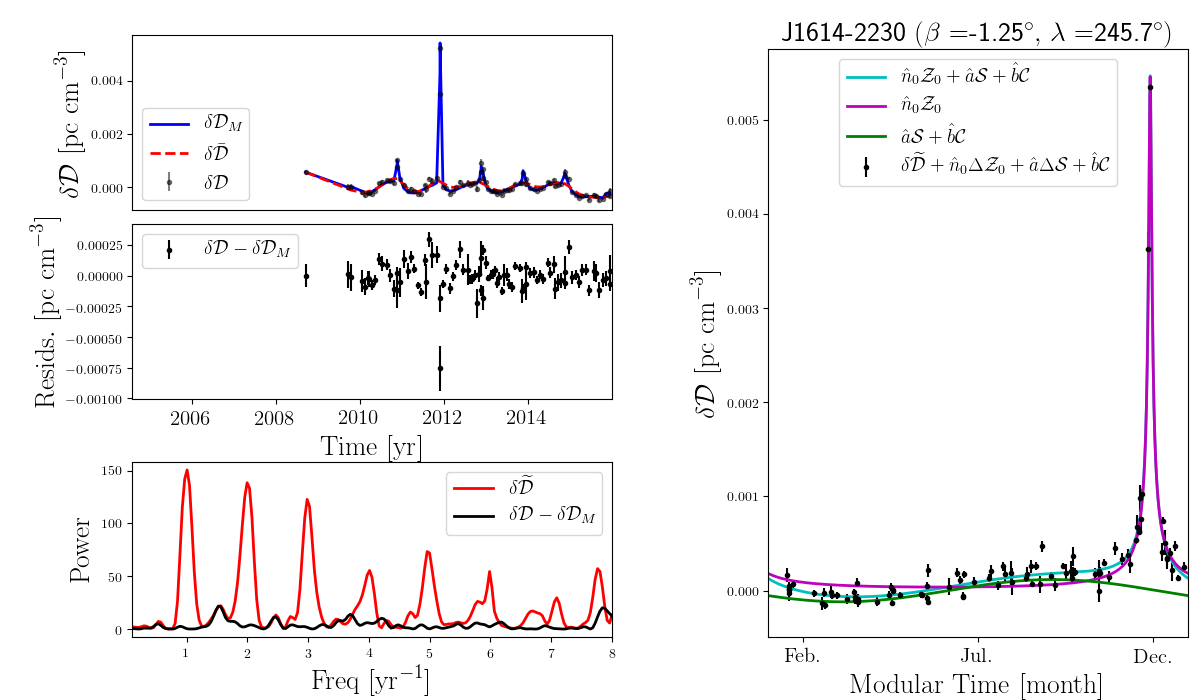}
\caption{\label{fig:J1614-2230}Modeling results as applied to PSR J1614$-$2230. {\bf Top Left:} Measured variations in DM about a nominal value, $\delta{\cal D}$, are shown in black dots. The dashed red curve is a low-frequency approximation to the DM fluctuations, $\delta{\bar {\cal D}}$, the result of convolving $\delta{\cal D}$ with a Gaussian (see Eqs. \ref{eq:smoothing} and \ref{eq:weights}). The blue curve represents the best fit model, $\delta{\cal D}_M$, as described in Eq. \ref{eq:fullModel}. {\bf Middle Left:} Model residuals, i.e. $\delta{\cal D}-\delta{\cal D}_M$. {\bf Bottom Left:} Unnormalized Lomb-Scargle periodograms of the high-frequency component of the DM fluctuations, $\delta\widetilde{\cal D}$, and the residuals, $\delta{\cal D}-\delta{\cal D}_M$. {\bf Right:} The black dots are the high-pass filtered DM data plotted modulo 1 yr; we have added in the basis corrections $\Delta{\cal Z}_0$, $\Delta{\cal S}$, and $\Delta{\cal C}$ scaled by the best-fit coefficients $\hat{n}_0$, $\hat{a}$, and $\hat{b}$ to compare the filtered DM data and the basis functions ${\cal Z}_0$, ${\cal S}$, and ${\cal C}$ (see the discussion following Eq. \ref{eq:filteredData}). The green (magenta) curve is the best-fit contribution from the ISM (SW). The cyan curve is the sum of the magenta and green curves.}
\end{center}
\end{figure*}

\begin{figure*}
\begin{center}
\includegraphics[scale=.5]{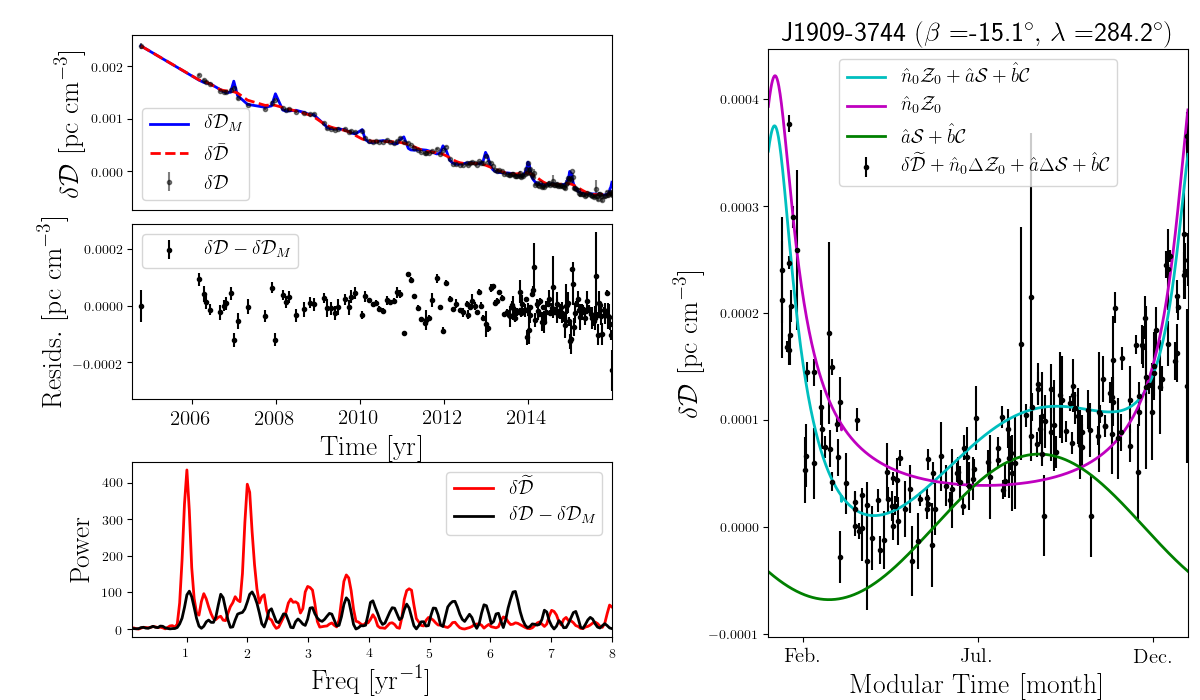}
\caption{\label{fig:J1909-3744}As in Figure \ref{fig:J1614-2230}, but for PSR J1909$-$3744.}
\end{center}
\end{figure*}

In Figures \ref{fig:J1614-2230} and \ref{fig:J1909-3744}, we show the results of our modeling in detail for two pulsars: J1614$-$2230, the NANOGrav pulsar closest to the ecliptic; and J1909$-$3744, arguably the single best timed of the NANOGrav pulsars \citep{lmc+18}. 

The model residuals for J1614$-$2230 in Figure \ref{fig:J1614-2230} show no obvious structure. The largest outlier is associated with the observation taken \emph{second} closest to the Sun. It is possible that we can not satisfactorily model both of the two closest observations to the Sun for this pulsar without considering non-inverse-square components of the SW. It is also possible that a discrete event such as a coronal mass ejection influenced this individual measurement as was the case in \citet{hsd+16}; for observations taken so close to the Sun, such considerations may become important. In the periodogram\footnote[6]{We have used the un-normalized Lomb-Scargle periodogram as described in Equation 12 of \citet{vi15} and implemented in Astropy \citep{astropy}.}, it is clear that the many harmonics of 1 yr$^{-1}$ present in the DM fluctuations of this pulsar have been successfully mitigated. In the right panel, one can see that the annual fluctuations in DM from the ISM are substantially sub-dominant to the fluctuations from the SW.

The model residuals for J1909$-$3744 in Figure \ref{fig:J1909-3744} show much more structure than those of J1614$-$2230 in Figure \ref{fig:J1614-2230}. It is possible that this is unmodeled influence of the ISM. J1909$-$3744 has a high proper motion \citep{mnf+16}, which we argue tends to reduce the periodic content of ISM-induced DM fluctuations. However, the right panel of Figure \ref{fig:J1909-3744} shows a clear ``shoulder'' in this pulsar's annual DM fluctuations that is well fit by the green curve describing an annual sinusoid from the ISM. This may indicate a screen of ISM material between Earth and J1909$-$3744 that is very close to the solar system. Additionally, Figure \ref{fig:Zl} shows that for pulsars further from the ecliptic, though the perturbation to DM from the SW is smaller in amplitude, it is spread out over a bigger percentage of orbital phase. It is possible that our observing cadence is high enough and the DM measurement precision for J1909$-$3744 is good enough that unmodeled latitudinal structure in the SW is showing up in the model residuals. 

\begin{figure}
\begin{center}
\includegraphics[scale=.5]{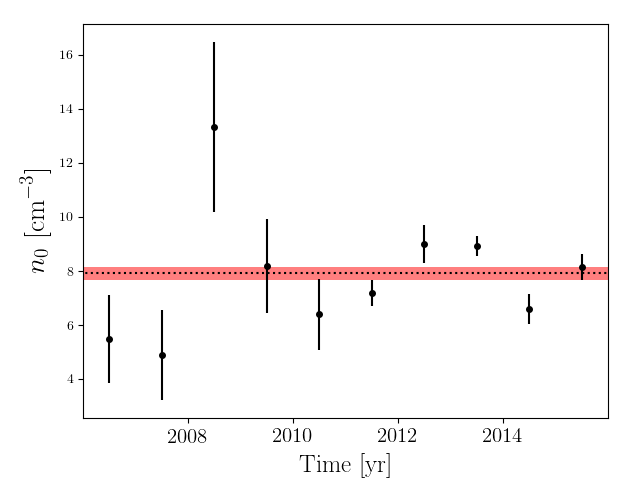}
\caption{\label{fig:timeResolved}Best-fit values of the solar wind electron density at 1 A.U., $n_0$, when it is allowed to vary from year to year. We show results for only the final 10 years of the 11-yr data set because the best-fit values for $n_0$ from the first approximately 1.4 yr of data are within 2-$\sigma$ of zero, effectively a non-detection. The black dotted line shows the best-fit value of $n_0$ when we assume it to be constant throughout our data set; the red shaded region indicates the 1-$\sigma$ uncertainties on that value.}
\end{center}
\end{figure}

When we relax the requirement that $n_0$ be constant throughout our data set and allow for it to be piecewise constant as discussed at the end of the previous section, we get the results shown in Figure \ref{fig:timeResolved}. Although we allowed $n_0$ to take on a different value in each of the 12 years that the 11-yr data set spills into, we only show the results for the final 10 years. The first value of $n_0$ we exclude from the plot is $1.9\pm4.6$ cm$^{-3}$: only 18 of the 3321 DM measurements we used fall into the span of times constraining this first value of $n_0$, only 11 of our 45 pulsars have data going that far back in time, and all but two of those pulsars are more than 10 degrees from the ecliptic. The second value we exclude from the plot is $-2.5\pm1.7$ cm$^{-3}$. Our fitting procedure does not restrict $n_0$ to positive values, but only positive values are physically meaningful. Only 62 DM measurements from just 18 pulsars are used to determine this value of $n_0$, but just 11 of those DM measurements are from the three pulsars within 10 degrees of the ecliptic. But the main issue with this second excluded value is that it is centered on a year where the Green Bank Telescope was off line for much of the year, leaving almost a year-long gap in our observations of many pulsars. In the final 10 years of results we show, there is clear improvement in measurement precision over time as more pulsars were added to the NANOGrav timing program and hardware at our telescopes was upgraded. Variations in the wind density about the value we get when it is assumed to be constant (the dotted black line with 1-$\sigma$ uncertainty indicated in red) are noise-like, indicating that there is little evidence in our data for time evolution in $n_0$.      

Table 2 summarizes our results by presenting $\chi^2$ values for various steps of our modeling on a pulsar by pulsar basis. If ${\cal R}$ are the residuals of a particular model, $\chi^2={\cal R}^{\rm T}\Xi^{-1}{\cal R}$. We have divided the $\chi^2$ values by $N_{\rm obs}$, the number of DM observations for a pulsar. This is approximately equal to the number of degrees of freedom for that pulsar, but straightforwardly determining the number of degrees of freedom for a particular pulsar is complicated by the nature of our modeling; the filtering we do is not equivalent to fitting out a parameterized model, and some of our fit parameters affect pulsars individually (an annual sine and cosine per pulsar) while some parameters affect all pulsars by varying amounts depending on the pulsar's ecliptic latitude (a constant or piecewise constant $n_0$). For instances when we have fit out an annual sine and cosine per pulsar, we divide by $(N_{\rm obs}-2)$. 

In Table 2, we include the sum of the $N_{\rm obs}$ and all $\chi^2$ columns. These sums show that high-pass filtering dramatically reduces the global $\chi^2$, an annual sine and cosine per pulsar plus a constant $n_0$ model further substantially reduces the global $\chi^2$, and a time variable $n_0$ reduces the global $\chi^2$ marginally further. This picture of steady fit improvement at each subsequent step of our modeling is complicated when tested on a pulsar by pulsar basis. For J0030$+$0451, the pulsar second closest to the ecliptic in our sample, the $\chi^2$ value improves at each step. For J1614$-$2230, the pulsar closest to the ecliptic in our sample, the constant $n_0$ model dramatically improves the $\chi^2$ value as compared to the high-pass filtered case, but the $\chi^2$ value is made marginally worse by allowing $n_0$ to vary. Notably, these pulsars probe different hemispheres of the SW and we may just be seeing that time evolution without latitudinal variation is insufficient for modeling our most near-ecliptic pulsars. The high values of $\chi^2$ for many pulsars indicate that our DM measurement uncertainties are very small and that there is structure left over in our DM time series from still-unmodeled phenomenology in the ISM and SW.            

\begin{deluxetable*}{l|c|c|c|c|c|c}
\tablewidth{0pc}
\tablecaption{$\chi^2$ values for different models.}
\tablehead{
	\colhead{PSR}&
    \colhead{$\beta$}&
	\colhead{$N_{\rm obs}$}&
    \colhead{$\chi^2/N_{\rm obs}$}&
    \colhead{$\chi^2/N_{\rm obs}$}&
    \colhead{$\chi^2/(N_{\rm obs}-2)$}&
    \colhead{$\chi^2/(N_{\rm obs}-2)$}\\
    \colhead{}&
    \colhead{[deg]}&
    \colhead{}&
    \colhead{\emph{No Model}}&
    \colhead{\emph{High Pass}}&
    \colhead{\emph{Fix $n_0$}}&
    \colhead{\emph{Vary $n_0$}}
	}
\startdata
J0023$+$0923    &       6.3     &       50      &       2527.9  &       1615.8  &       666.0   &       643.1\\
J0030$+$0451    &       1.4     &       102     &       738.6   &       715.0   &       107.5   &       76.8\\
J0340$+$4130    &       21.3    &       56      &       119.8   &       2.1     &       1.8     &       1.8\\
J0613$-$0200    &       $-$25.4 &       121     &       1281.7  &       20.3    &       12.0    &       12.2\\
J0636$+$5128    &       28.2    &       26      &       24.1    &       9.5     &       5.9     &       6.0\\
J0645$+$5158    &       28.8    &       61      &       18.2    &       11.0    &       10.4    &       10.4\\
J0740$+$6620    &       44.1    &       26      &       7.0     &       4.6     &       4.8     &       4.8\\
J0931$-$1902    &       $-$31.7 &       39      &       3.8     &       2.2     &       2.0     &       2.0\\
J1012$+$5307    &       38.7    &       123     &       45.3    &       3.6     &       3.5     &       3.5\\
J1024$-$0719    &       $-$16.0 &       82      &       81.4    &       4.0     &       3.0     &       3.1\\
J1125$+$7819    &       62.4    &       25      &       372.8   &       106.0   &       95.7    &       95.8\\
J1453$+$1902    &       33.9    &       22      &       5.4     &       6.7     &       6.9     &       6.9\\
J1455$-$3330    &       $-$16.0 &       108     &       21.6    &       9.5     &       8.3     &       8.6\\
J1600$-$3053    &       $-$10.0 &       106     &       612.2   &       15.5    &       9.5     &       9.7\\
J1614$-$2230    &       $-$1.2  &       92      &       311.0   &       260.2   &       6.6     &       5.8\\
J1640$+$2224    &       44.0    &       111     &       1788.6  &       84.3    &       72.6    &       77.5\\
J1643$-$1224    &       9.7     &       122     &       3061.2  &       62.1    &       57.5    &       56.1\\
J1713$+$0747    &       30.7    &       198     &       85.7    &       26.7    &       26.1    &       20.4\\
J1738$+$0333    &       26.8    &       54      &       181.3   &       9.5     &       9.0     &       9.1\\
J1741$+$1351    &       37.2    &       59      &       506.0   &       38.3    &       35.4    &       35.7\\
J1744$-$1134    &       11.8    &       116     &       300.7   &       53.9    &       44.6    &       47.8\\
J1747$-$4036    &       $-$17.2 &       54      &       138.1   &       11.6    &       11.1    &       11.1\\
J1832$-$0836    &       14.5    &       39      &       503.2   &       10.5    &       4.4     &       4.6\\
J1853$+$1303    &       35.7    &       53      &       253.8   &       292.1   &       292.6   &       292.1\\
B1855$+$09      &       32.3    &       101     &       2298.0  &       12.2    &       13.4    &       13.9\\
J1903$+$0327    &       25.9    &       60      &       1961.5  &       8.0     &       6.1     &       6.1\\
J1909$-$3744    &       $-$15.1 &       166     &       4013.2  &       27.5    &       15.8    &       15.0\\
J1910$+$1256    &       35.1    &       67      &       25.6    &       6.2     &       5.9     &       5.8\\
J1911$+$1347    &       35.8    &       25      &       165.5   &       5.2     &       1.5     &       1.6\\
J1918$-$0642    &       15.3    &       117     &       374.7   &       5.0     &       3.4     &       3.1\\
J1923$+$2515    &       46.6    &       48      &       51.1    &       25.8    &       20.3    &       20.9\\
B1937$+$21      &       42.2    &       165     &       22133.8 &       1118.3  &       1013.0  &       1025.7\\
J1944$+$0907    &       29.8    &       53      &       2376.1  &       36.4    &       29.7    &       29.6\\
B1953$+$29      &       48.6    &       47      &       363.7   &       78.2    &       79.4    &       79.3\\
J2010$-$1323    &       6.4     &       88      &       139.1   &       29.1    &       5.2     &       5.4\\
J2017$+$0603    &       25.0    &       49      &       43.2    &       16.9    &       7.6     &       7.8\\
J2033$+$1734    &       35.0    &       23      &       28.0    &       19.9    &       22.1    &       22.5\\
J2043$+$1711    &       33.9    &       65      &       652.2   &       271.5   &       227.2   &       225.9\\
J2145$-$0750    &       5.3     &       107     &       181.9   &       49.7    &       64.5    &       60.5\\
J2214$+$3000    &       37.7    &       53      &       32.5    &       38.0    &       32.1    &       32.2\\
J2229$+$2643    &       33.2    &       21      &       3.0     &       2.2     &       2.4     &       2.4\\
J2234$+$0611    &       14.0    &       23      &       17.8    &       5.8     &       5.5     &       5.4\\
J2234$+$0944    &       17.3    &       29      &       80.9    &       44.6    &       37.5    &       37.1\\
J2302$+$4442    &       45.6    &       58      &       23.6    &       8.4     &       8.6     &       8.7\\
J2317$+$1439    &       17.6    &       111     &       133009.7        &       244.9   &       217.7   &       184.9\\\hline
SUM	&		&	3321	&	180966.6	&	5430.7	&	3318.0	&	3241.0	
\enddata
\end{deluxetable*}

\section{Prospects \& Concluding Remarks}

\begin{figure}
\begin{center}
\label{fig:population}
\includegraphics[scale=.57]{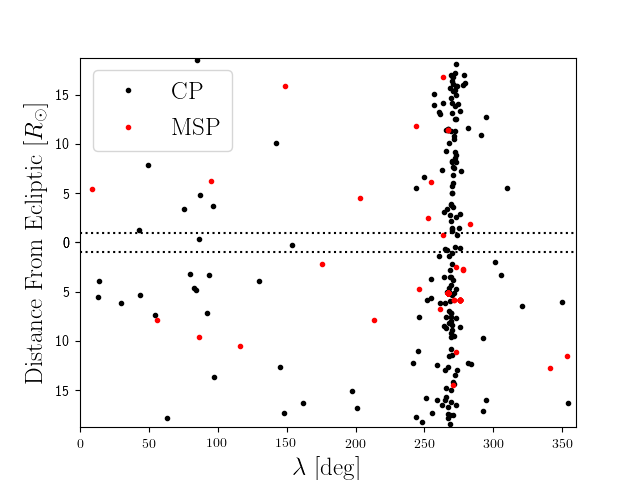}
\caption{\label{fig:population}All pulsars with $|\beta|\leq 5^\circ$ according to the ATNF pulsar catalog \citep{mht+05}. The lines of sight to these pulsars come within 20 solar radii of the Sun (as indicated by the $y$-axis). The dense vertical strip of pulsars near $\lambda=270^\circ$ are in the direction of the Galactic interior and approach the Sun around December or January of every year. The red dots represent millisecond pulsars (MSP) and the black dots represent canonical pulsars (CP). The horizontal dashed line represents the angular extent of the Sun.}
\end{center}
\end{figure}

The NANOGrav 11-yr data set is among the best collections of pulsar timing data in existence for looking for and studying nanohertz gravitational waves, rivaled only by similar data sets from the European Pulsar Timing Array \citep{kc13,dcl+16} and the Parkes Pulsar Timing Array \citep{h13,rhc+16}. But if one set out to observe pulsars for the purpose of investigating the SW rather than gravitational waves, the set of pulsars observed and the observing strategies employed would be quite different. 

\citet{tv18} recently presented low-frequency (approximately 100~MHz), high cadence (approximately weekly) observations of three pulsars within 9$^\circ$ of the ecliptic conducted with individual stations of the LOFAR telescope. Since dispersive timing delays scale as the inverse square of the radio frequency, variations in the electron content along the LOS lead to bigger and more precisely measurable timing fluctuations at these low radio frequencies. As Figure \ref{fig:Zl} shows, high cadence observations, particularly through solar conjunction, are necessary for probing latitudinal variations in the SW; higher than weekly cadence would be beneficial within approximately 10 days before and after solar conjunction. 

Figure \ref{fig:population} shows the entire known population of pulsars within 5$^\circ$ of the ecliptic. These pulsars come within 20~$R_\odot$ or less of the Sun when in solar conjunction, some of them being fully eclipsed by the Sun. For comparison, NASA's Parker Solar Probe will come within approximately 5 $R_\odot$ of the Sun's surface. A high-cadence, low frequency pulsar observing campaign through approximately December and January, when the bulk of the near-ecliptic pulsar population drifts behind the Sun, paired with the analysis techniques we have developed here, could map out the large scale structure of the SW and powerfully complement the \emph{in situ} capabilities of the Parker Solar Probe. New telescopes like the Canadian Hydrogen Intensity Mapping Experiment (CHIME) are well suited for this kind of high-cadence observational campaign.   

Additionally, pulsars are strongly linearly polarized and are thus very useful for probing the Sun's magnetic field \citep{bsv80,ych+12}. The type of many-pulsar analysis we have developed in this work could be straightforwardly extended to an analysis of rotation measures to make unprecedented inferences about the large scale configuration of the Sun's magnetic field. A high cadence, low frequency observing campaign would also be ideal for this application.\newline     

\emph{Author contributions}: D.R.M. wrote this manuscript and developed the techniques it describes. J.M.C. made code for simulating screens of turbulent material in the interstellar medium. D.J.N. conducted a preliminary analysis of the annual component of dispersion measure variations. J.M.C., C.M.F.M., D.J.N., M.T.L, M.A.M, and S.C. reviewed and substantially improved this manuscript. Z.A., K.C., P.B.D., M.E.D., T.D., J.A.E., R.D.F., E.C.F., E.F., P.A.G., G.J., M.L.J., M.T.L., L.L., D.R.L., R.S.L., M.A.M., C.N., D.J.N., T.T.P., S.M.R., P.S.R., R.S., I.H.S., K.S., J.K.S., and W.Z. contributed to the development of the 11-yr data set, as detailed in \citet{abb+18_a}.

\begin{acknowledgments}
The NANOGrav project receives support from National Science Foundation (NSF) Physics Frontiers Center award number 1430284. D.R.M. is a Jansky Fellow of the National Radio Astronomy Observatory (NRAO). NRAO is a facility of the NSF operated under cooperative agreement by Associated Universities, Inc. The Flatiron Institute is supported by the Simons Foundation. Pulsar research at UBC is supported by an NSERC Discovery Grant and by the Canadian Institute for Advanced Research.
\end{acknowledgments}

\bibliographystyle{aasjournal}
\bibliography{solarWind}

\begin{thebibliography}{}
\expandafter\ifx\csname natexlab\endcsname\relax\def\natexlab#1{#1}\fi
\providecommand{\url}[1]{\href{#1}{#1}}

\bibitem[{{Archibald} {et~al.}(2018){Archibald}, {Gusinskaia}, {Hessels},
  {Deller}, {Kaplan}, {Lorimer}, {Lynch}, {Ransom}, \& {Stairs}}]{agh+18}
{Archibald}, A.~M., {Gusinskaia}, N.~V., {Hessels}, J.~W.~T., {et~al.} 2018,
  ArXiv e-prints, arXiv:1807.02059

\bibitem[{{Arzoumanian} {et~al.}(2018{\natexlab{a}}){Arzoumanian}, {Brazier},
  {Burke-Spolaor}, {Chamberlin}, {Chatterjee}, {Christy}, {Cordes}, {Cornish},
  {Crawford}, {Thankful Cromartie}, {Crowter}, {DeCesar}, {Demorest}, {Dolch},
  {Ellis}, {Ferdman}, {Ferrara}, {Fonseca}, {Garver-Daniels}, {Gentile},
  {Halmrast}, {Huerta}, {Jenet}, {Jessup}, {Jones}, {Jones}, {Kaplan}, {Lam},
  {Lazio}, {Levin}, {Lommen}, {Lorimer}, {Luo}, {Lynch}, {Madison}, {Matthews},
  {McLaughlin}, {McWilliams}, {Mingarelli}, {Ng}, {Nice}, {Pennucci}, {Ransom},
  {Ray}, {Siemens}, {Simon}, {Spiewak}, {Stairs}, {Stinebring}, {Stovall},
  {Swiggum}, {Taylor}, {Vallisneri}, {van Haasteren}, {Vigeland}, {Zhu}, \&
  {The NANOGrav Collaboration}}]{abb+18_a}
{Arzoumanian}, Z., {Brazier}, A., {Burke-Spolaor}, S., {et~al.}
  2018{\natexlab{a}}, \apjs, 235, 37

\bibitem[{{Arzoumanian} {et~al.}(2018{\natexlab{b}}){Arzoumanian}, {Baker},
  {Brazier}, {Burke-Spolaor}, {Chamberlin}, {Chatterjee}, {Christy}, {Cordes},
  {Cornish}, {Crawford}, {Thankful Cromartie}, {Crowter}, {DeCesar},
  {Demorest}, {Dolch}, {Ellis}, {Ferdman}, {Ferrara}, {Folkner}, {Fonseca},
  {Garver-Daniels}, {Gentile}, {Haas}, {Hazboun}, {Huerta}, {Islo}, {Jones},
  {Jones}, {Kaplan}, {Kaspi}, {Lam}, {Lazio}, {Levin}, {Lommen}, {Lorimer},
  {Luo}, {Lynch}, {Madison}, {McLaughlin}, {McWilliams}, {Mingarelli}, {Ng},
  {Nice}, {Park}, {Pennucci}, {Pol}, {Ransom}, {Ray}, {Rasskazov}, {Siemens},
  {Simon}, {Spiewak}, {Stairs}, {Stinebring}, {Stovall}, {Swiggum}, {Taylor},
  {Vallisneri}, {van Haasteren}, {Vigeland}, {Zhu}, \& {The NANOGrav
  Collaboration}}]{abb+18_b}
{Arzoumanian}, Z., {Baker}, P.~T., {Brazier}, A., {et~al.} 2018{\natexlab{b}},
  \apj, 859, 47

\bibitem[{{Bird} {et~al.}(1980){Bird}, {Schruefer}, {Volland}, \&
  {Sieber}}]{bsv80}
{Bird}, M.~K., {Schruefer}, E., {Volland}, H., \& {Sieber}, W. 1980, \nat, 283,
  459

\bibitem[{{Coles} {et~al.}(2015){Coles}, {Kerr}, {Shannon}, {Hobbs},
  {Manchester}, {You}, {Bailes}, {Bhat}, {Burke-Spolaor}, {Dai}, {Keith},
  {Levin}, {Os{\l}owski}, {Ravi}, {Reardon}, {Toomey}, {van Straten}, {Wang},
  {Wen}, \& {Zhu}}]{cks+15}
{Coles}, W.~A., {Kerr}, M., {Shannon}, R.~M., {et~al.} 2015, \apj, 808, 113

\bibitem[{{Cordes} {et~al.}(2016){Cordes}, {Shannon}, \& {Stinebring}}]{css16}
{Cordes}, J.~M., {Shannon}, R.~M., \& {Stinebring}, D.~R. 2016, \apj, 817, 16

\bibitem[{{Counselman} \& {Rankin}(1972)}]{cr72}
{Counselman}, III, C.~C., \& {Rankin}, J.~M. 1972, \apj, 175, 843

\bibitem[{{Counselman} \& {Shapiro}(1968)}]{cs68}
{Counselman}, III, C.~C., \& {Shapiro}, I.~I. 1968, Science, 162, 352

\bibitem[{{Desvignes} {et~al.}(2016){Desvignes}, {Caballero}, {Lentati},
  {Verbiest}, {Champion}, {Stappers}, {Janssen}, {Lazarus}, {Os{\l}owski},
  {Babak}, {Bassa}, {Brem}, {Burgay}, {Cognard}, {Gair}, {Graikou},
  {Guillemot}, {Hessels}, {Jessner}, {Jordan}, {Karuppusamy}, {Kramer},
  {Lassus}, {Lazaridis}, {Lee}, {Liu}, {Lyne}, {McKee}, {Mingarelli},
  {Perrodin}, {Petiteau}, {Possenti}, {Purver}, {Rosado}, {Sanidas}, {Sesana},
  {Shaifullah}, {Smits}, {Taylor}, {Theureau}, {Tiburzi}, {van Haasteren}, \&
  {Vecchio}}]{dcl+16}
{Desvignes}, G., {Caballero}, R.~N., {Lentati}, L., {et~al.} 2016, \mnras, 458,
  3341

\bibitem[{{DuPlain} {et~al.}(2008){DuPlain}, {Ransom}, {Demorest}, {Brandt},
  {Ford}, \& {Shelton}}]{drd+08}
{DuPlain}, R., {Ransom}, S., {Demorest}, P., {et~al.} 2008, in \procspie, Vol.
  7019, Advanced Software and Control for Astronomy II, 70191D

\bibitem[{{Edwards} {et~al.}(2006){Edwards}, {Hobbs}, \& {Manchester}}]{ehm06}
{Edwards}, R.~T., {Hobbs}, G.~B., \& {Manchester}, R.~N. 2006, \mnras, 372,
  1549

\bibitem[{{Goldstein} \& {Meisel}(1969)}]{gm69}
{Goldstein}, S.~J., \& {Meisel}, D.~D. 1969, \nat, 224, 349

\bibitem[{{Hobbs}(2013)}]{h13}
{Hobbs}, G. 2013, Classical and Quantum Gravity, 30, 224007

\bibitem[{{Hollweg}(1968)}]{h68}
{Hollweg}, J.~V. 1968, \nat, 220, 771

\bibitem[{{Howard} {et~al.}(2016){Howard}, {Stovall}, {Dowell}, {Taylor}, \&
  {White}}]{hsd+16}
{Howard}, T.~A., {Stovall}, K., {Dowell}, J., {Taylor}, G.~B., \& {White},
  S.~M. 2016, \apj, 831, 208

\bibitem[{{Issautier} {et~al.}(2008){Issautier}, {Le Chat}, {Meyer-Vernet},
  {Moncuquet}, {Hoang}, {MacDowall}, \& {McComas}}]{ilm+08}
{Issautier}, K., {Le Chat}, G., {Meyer-Vernet}, N., {et~al.} 2008, \grl, 35,
  L19101

\bibitem[{{Issautier} {et~al.}(1997){Issautier}, {Meyer-Vernet}, {Moncuquet},
  \& {Hoang}}]{imm+97}
{Issautier}, K., {Meyer-Vernet}, N., {Moncuquet}, M., \& {Hoang}, S. 1997,
  \solphys, 172, 335

\bibitem[{{Issautier} {et~al.}(1998){Issautier}, {Meyer-Vernet}, {Moncuquet},
  \& {Hoang}}]{imm+98}
---. 1998, \jgr, 103, 1969

\bibitem[{{Issautier} {et~al.}(2004){Issautier}, {Moncuquet}, \&
  {Hoang}}]{imh04}
{Issautier}, K., {Moncuquet}, M., \& {Hoang}, S. 2004, \solphys, 221, 351

\bibitem[{{Jones} {et~al.}(2017){Jones}, {McLaughlin}, {Lam}, {Cordes},
  {Levin}, {Chatterjee}, {Arzoumanian}, {Crowter}, {Demorest}, {Dolch},
  {Ellis}, {Ferdman}, {Fonseca}, {Gonzalez}, {Jones}, {Lazio}, {Nice},
  {Pennucci}, {Ransom}, {Stinebring}, {Stairs}, {Stovall}, {Swiggum}, \&
  {Zhu}}]{jml+17}
{Jones}, M.~L., {McLaughlin}, M.~A., {Lam}, M.~T., {et~al.} 2017, \apj, 841,
  125

\bibitem[{{Kramer} \& {Champion}(2013)}]{kc13}
{Kramer}, M., \& {Champion}, D.~J. 2013, Classical and Quantum Gravity, 30,
  224009

\bibitem[{{Lam} {et~al.}(2015){Lam}, {Cordes}, {Chatterjee}, \&
  {Dolch}}]{lcc+15}
{Lam}, M.~T., {Cordes}, J.~M., {Chatterjee}, S., \& {Dolch}, T. 2015, \apj,
  801, 130

\bibitem[{{Lam} {et~al.}(2018{\natexlab{a}}){Lam}, {McLaughlin}, {Cordes},
  {Chatterjee}, \& {Lazio}}]{lmc+18}
{Lam}, M.~T., {McLaughlin}, M.~A., {Cordes}, J.~M., {Chatterjee}, S., \&
  {Lazio}, T.~J.~W. 2018{\natexlab{a}}, \apj, 861, 12

\bibitem[{{Lam} {et~al.}(2016){Lam}, {Cordes}, {Chatterjee}, {Arzoumanian},
  {Crowter}, {Demorest}, {Dolch}, {Ellis}, {Ferdman}, {Fonseca}, {Gonzalez},
  {Jones}, {Jones}, {Levin}, {Madison}, {McLaughlin}, {Nice}, {Pennucci},
  {Ransom}, {Siemens}, {Stairs}, {Stovall}, {Swiggum}, \& {Zhu}}]{lcc+16_1}
{Lam}, M.~T., {Cordes}, J.~M., {Chatterjee}, S., {et~al.} 2016, \apj, 819, 155

\bibitem[{{Lam} {et~al.}(2017){Lam}, {Cordes}, {Chatterjee}, {Arzoumanian},
  {Crowter}, {Demorest}, {Dolch}, {Ellis}, {Ferdman}, {Fonseca}, {Gonzalez},
  {Jones}, {Jones}, {Levin}, {Madison}, {McLaughlin}, {Nice}, {Pennucci},
  {Ransom}, {Shannon}, {Siemens}, {Stairs}, {Stovall}, {Swiggum}, \&
  {Zhu}}]{lcc+17}
---. 2017, \apj, 834, 35

\bibitem[{{Lam} {et~al.}(2018{\natexlab{b}}){Lam}, {Ellis}, {Grillo}, {Jones},
  {Hazboun}, {Brook}, {Turner}, {Chatterjee}, {Cordes}, {Lazio}, {DeCesar},
  {Arzoumanian}, {Blumer}, {Cromartie}, {Demorest}, {Dolch}, {Ferdman},
  {Ferrara}, {Fonseca}, {Garver-Daniels}, {Gentile}, {Gupta}, {Lorimer},
  {Lynch}, {Madison}, {McLaughlin}, {Ng}, {Nice}, {Pennucci}, {Ransom},
  {Spiewak}, {Stairs}, {Stinebring}, {Stovall}, {Swiggum}, {Vigeland}, \&
  {Zhu}}]{leg+18}
{Lam}, M.~T., {Ellis}, J.~A., {Grillo}, G., {et~al.} 2018{\natexlab{b}}, \apj,
  861, 132

\bibitem[{{Leblanc} {et~al.}(1998){Leblanc}, {Dulk}, \& {Bougeret}}]{ldb98}
{Leblanc}, Y., {Dulk}, G.~A., \& {Bougeret}, J.-L. 1998, \solphys, 183, 165

\bibitem[{{Levin} {et~al.}(2016){Levin}, {McLaughlin}, {Jones}, {Cordes},
  {Stinebring}, {Chatterjee}, {Dolch}, {Lam}, {Lazio}, {Palliyaguru},
  {Arzoumanian}, {Crowter}, {Demorest}, {Ellis}, {Ferdman}, {Fonseca},
  {Gonzalez}, {Jones}, {Nice}, {Pennucci}, {Ransom}, {Stairs}, {Stovall},
  {Swiggum}, \& {Zhu}}]{lmj+16}
{Levin}, L., {McLaughlin}, M.~A., {Jones}, G., {et~al.} 2016, \apj, 818, 166

\bibitem[{{Lommen} {et~al.}(2006){Lommen}, {Kipphorn}, {Nice}, {Splaver},
  {Stairs}, \& {Backer}}]{lkn+06}
{Lommen}, A.~N., {Kipphorn}, R.~A., {Nice}, D.~J., {et~al.} 2006, \apj, 642,
  1012

\bibitem[{{Manchester} {et~al.}(2005){Manchester}, {Hobbs}, {Teoh}, \&
  {Hobbs}}]{mht+05}
{Manchester}, R.~N., {Hobbs}, G.~B., {Teoh}, A., \& {Hobbs}, M. 2005, \aj, 129,
  1993

\bibitem[{{Matthews} {et~al.}(2016){Matthews}, {Nice}, {Fonseca},
  {Arzoumanian}, {Crowter}, {Demorest}, {Dolch}, {Ellis}, {Ferdman},
  {Gonzalez}, {Jones}, {Jones}, {Lam}, {Levin}, {McLaughlin}, {Pennucci},
  {Ransom}, {Stairs}, {Stovall}, {Swiggum}, \& {Zhu}}]{mnf+16}
{Matthews}, A.~M., {Nice}, D.~J., {Fonseca}, E., {et~al.} 2016, \apj, 818, 92

\bibitem[{{Nice} {et~al.}(2015){Nice}, {Demorest}, {Stairs}, {Manchester},
  {Taylor}, {Peters}, {Weisberg}, {Irwin}, {Wex}, \& {Huang}}]{nds+15}
{Nice}, D., {Demorest}, P., {Stairs}, I., {et~al.} 2015, {Tempo: Pulsar timing
  data analysis}, Astrophysics Source Code Library, , , ascl:1509.002

\bibitem[{{Niu} {et~al.}(2017){Niu}, {Hobbs}, {Wang}, \& {Dai}}]{nhw+17}
{Niu}, Z.-X., {Hobbs}, G., {Wang}, J.-B., \& {Dai}, S. 2017, Research in
  Astronomy and Astrophysics, 17, 103

\bibitem[{{Ransom} {et~al.}(2009){Ransom}, {Demorest}, {Ford}, {McCullough},
  {Ray}, {DuPlain}, \& {Brandt}}]{rdf+09}
{Ransom}, S.~M., {Demorest}, P., {Ford}, J., {et~al.} 2009, in American
  Astronomical Society Meeting Abstracts, Vol. 214, American Astronomical
  Society Meeting Abstracts \#214, 605.08

\bibitem[{{Reardon} {et~al.}(2016){Reardon}, {Hobbs}, {Coles}, {Levin},
  {Keith}, {Bailes}, {Bhat}, {Burke-Spolaor}, {Dai}, {Kerr}, {Lasky},
  {Manchester}, {Os{\l}owski}, {Ravi}, {Shannon}, {van Straten}, {Toomey},
  {Wang}, {Wen}, {You}, \& {Zhu}}]{rhc+16}
{Reardon}, D.~J., {Hobbs}, G., {Coles}, W., {et~al.} 2016, \mnras, 455, 1751

\bibitem[{{Rickett}(1990)}]{r90}
{Rickett}, B.~J. 1990, \araa, 28, 561

\bibitem[{{Splaver} {et~al.}(2005){Splaver}, {Nice}, {Stairs}, {Lommen}, \&
  {Backer}}]{sns+05}
{Splaver}, E.~M., {Nice}, D.~J., {Stairs}, I.~H., {Lommen}, A.~N., \& {Backer},
  D.~C. 2005, \apj, 620, 405

\bibitem[{{The Astropy Collaboration} {et~al.}(2018){The Astropy
  Collaboration}, {Price-Whelan}, {Sip{\H o}cz}, {G{\"u}nther}, {Lim},
  {Crawford}, {Conseil}, {Shupe}, {Craig}, {Dencheva}, {Ginsburg},
  {VanderPlas}, {Bradley}, {P{\'e}rez-Su{\'a}rez}, {de Val-Borro}, {Aldcroft},
  {Cruz}, {Robitaille}, {Tollerud}, {Ardelean}, {Babej}, {Bachetti}, {Bakanov},
  {Bamford}, {Barentsen}, {Barmby}, {Baumbach}, {Berry}, {Biscani}, {Boquien},
  {Bostroem}, {Bouma}, {Brammer}, {Bray}, {Breytenbach}, {Buddelmeijer},
  {Burke}, {Calderone}, {Cano Rodr{\'{\i}}guez}, {Cara}, {Cardoso},
  {Cheedella}, {Copin}, {Crichton}, {D{\'A}vella}, {Deil}, {Depagne},
  {Dietrich}, {Donath}, {Droettboom}, {Earl}, {Erben}, {Fabbro}, {Ferreira},
  {Finethy}, {Fox}, {Garrison}, {Gibbons}, {Goldstein}, {Gommers}, {Greco},
  {Greenfield}, {Groener}, {Grollier}, {Hagen}, {Hirst}, {Homeier}, {Horton},
  {Hosseinzadeh}, {Hu}, {Hunkeler}, {Ivezi{\'c}}, {Jain}, {Jenness}, {Kanarek},
  {Kendrew}, {Kern}, {Kerzendorf}, {Khvalko}, {King}, {Kirkby}, {Kulkarni},
  {Kumar}, {Lee}, {Lenz}, {Littlefair}, {Ma}, {Macleod}, {Mastropietro},
  {McCully}, {Montagnac}, {Morris}, {Mueller}, {Mumford}, {Muna}, {Murphy},
  {Nelson}, {Nguyen}, {Ninan}, {N{\"o}the}, {Ogaz}, {Oh}, {Parejko}, {Parley},
  {Pascual}, {Patil}, {Patil}, {Plunkett}, {Prochaska}, {Rastogi}, {Reddy
  Janga}, {Sabater}, {Sakurikar}, {Seifert}, {Sherbert}, {Sherwood-Taylor},
  {Shih}, {Sick}, {Silbiger}, {Singanamalla}, {Singer}, {Sladen}, {Sooley},
  {Sornarajah}, {Streicher}, {Teuben}, {Thomas}, {Tremblay}, {Turner},
  {Terr{\'o}n}, {van Kerkwijk}, {de la Vega}, {Watkins}, {Weaver}, {Whitmore},
  {Woillez}, \& {Zabalza}}]{astropy}
{The Astropy Collaboration}, {Price-Whelan}, A.~M., {Sip{\H o}cz}, B.~M.,
  {et~al.} 2018, ArXiv e-prints, arXiv:1801.02634

\bibitem[{{Tiburzi} \& {Verbiest}(2018)}]{tv18}
{Tiburzi}, C., \& {Verbiest}, J. 2018, ArXiv e-prints, arXiv:1804.04040

\bibitem[{{van Haasteren} \& {Levin}(2013)}]{vl13}
{van Haasteren}, R., \& {Levin}, Y. 2013, \mnras, 428, 1147

\bibitem[{{VanderPlas} \& {Ivezi{\'c}}(2015)}]{vi15}
{VanderPlas}, J.~T., \& {Ivezi{\'c}}, {\v Z}. 2015, \apj, 812, 18

\bibitem[{{Vigeland} \& {Vallisneri}(2014)}]{vv14}
{Vigeland}, S.~J., \& {Vallisneri}, M. 2014, \mnras, 440, 1446

\bibitem[{{Wang} \& {Han}(2018)}]{wh18}
{Wang}, P.~F., \& {Han}, J.~L. 2018, \mnras, 479, 3393

\bibitem[{{You} {et~al.}(2012){You}, {Coles}, {Hobbs}, \&
  {Manchester}}]{ych+12}
{You}, X.~P., {Coles}, W.~A., {Hobbs}, G.~B., \& {Manchester}, R.~N. 2012,
  \mnras, 422, 1160

\bibitem[{{You} {et~al.}(2007){You}, {Hobbs}, {Coles}, {Manchester}, \&
  {Han}}]{yhc+07}
{You}, X.~P., {Hobbs}, G.~B., {Coles}, W.~A., {Manchester}, R.~N., \& {Han},
  J.~L. 2007, \apj, 671, 907

\end{thebibliography}
\end{document}